\documentclass{article}

\PassOptionsToPackage{numbers,compress}{natbib}
\usepackage[preprint]{neurips_2026}
\usepackage[utf8]{inputenc}
\usepackage[T1]{fontenc}
\usepackage{hyperref}
\usepackage{url}
\usepackage{booktabs}
\usepackage{amsfonts}
\usepackage{nicefrac}
\usepackage{microtype}
\usepackage{xcolor}

\usepackage{enumitem}

\usepackage{graphicx}
\usepackage{subcaption}
\usepackage{tabularx}
\usepackage{multirow}
\usepackage{amsmath}
\usepackage{amssymb}
\usepackage{mathtools}
\usepackage{amsthm}
\usepackage[capitalize,noabbrev]{cleveref}
\usepackage[binary-units]{siunitx}
\usepackage{listings}
\usepackage[most]{tcolorbox}
\tcbuselibrary{breakable}
\tcbset{
  rqbox/.style={
    enhanced,
    colback=gray!6,
    colframe=black!40,
    arc=3pt,
    boxrule=0.6pt,
    left=12pt, right=12pt, top=6pt, bottom=6pt,
    width=\linewidth,
    before skip=8pt,
    after skip=8pt,
  }
}

\theoremstyle{plain}

\theoremstyle{definition}

\theoremstyle{remark}

\usepackage[disable,textsize=tiny]{todonotes}

\newif\ifcomments

\commentsfalse

\ifcomments
\newcommand{\cd}[1]{\textcolor{blue}{[CD: #1]}}
\newcommand{\as}[1]{\textcolor{green}{[AS: #1]}}
\newcommand{\wh}[1]{\textcolor{red}{[MW: #1]}}

\else
\newcommand{\cd}[1]{}
\newcommand{\as}[1]{}
\newcommand{\wh}[1]{}
\fi

\newif\ifdraft
\draftfalse  

\providecommand{\wh}[1]{}
\providecommand{\cd}[1]{}
\providecommand{\as}[1]{}

\ifdraft
  \renewcommand{\wh}[1]{\textcolor{red}{\textbf{WH:} #1}}
  \renewcommand{\cd}[1]{\textcolor{blue}{\textbf{CD:} #1}}
  \renewcommand{\as}[1]{\textcolor{brown}{\textbf{AS:} #1}}
\else
  \renewcommand{\wh}[1]{}
  \renewcommand{\cd}[1]{}
  \renewcommand{\as}[1]{}
\fi

\newcommand{\rao}{SDE}
\newcommand{\dcd}{DSDE}

\newcommand{\tpass}{partial\_pass@1}

\newcommand{\raodcd}{distance-aware uncertainty metrics}

\title{Using Semantic Distance to Estimate Uncertainty in LLM-Based Code Generation}

\author{%
  \href{https://openreview.net/profile?id=~Weilin_He1}{Weilin He} \\
  University of Bristol \\
  \texttt{te23071@bristol.ac.uk} \\
  \And
  \href{https://openreview.net/profile?id=~Arindam_Sharma1}{Arindam Sharma} \\
  University of Bristol \\
  \texttt{arindam.sharma@bristol.ac.uk} \\
  \And
  \href{https://openreview.net/profile?id=~Cristina_David1}{Cristina David} \\
  University of Bristol \\
  \texttt{cristina.david@bristol.ac.uk} \\
}

\begin{document}

\maketitle

\begin{abstract}
  
LLMs show strong performance in code generation, but their outputs lack
correctness guarantees. Sample-based uncertainty estimators
address this by generating multiple candidate programs and measuring
their disagreement. However, existing estimators make different design
choices about how behaviours are identified, aggregated, referenced
and compared, making them difficult to assess. We therefore
first introduce a taxonomy that disentangles these choices and reveals
a missing design point: \emph{semantic distance-aware uncertainty estimation}, which measures
not only whether sampled programs disagree, but how severely their
execution behaviours differ.
Across
LiveCodeBench, MBPP, HumanEval-X and BigCodeBench, spanning Python, Java and C++, our metrics provide
strong proxies for correctness, and consistently outperform state-of-the-art
sample-based baselines across both closed-source models (GPT-3.5-Turbo, GPT-4o-mini, Gemini-2.5-Flash-Lite, Claude Opus 4.5) and an open-source model (DeepSeek-Coder-V2). The method is practical: it requires neither
model internals nor LLM-as-judge calls, remains robust across models,
languages, sampling temperatures and fuzzing settings, and reduces
runtime by approximately $48$--$79\%$ relative to existing baselines.

\end{abstract}

\section{Introduction}

The adoption of Large Language Models (LLMs) in software engineering
has been rapid, with the volume of LLM-generated code projected to
increase further~\citep{debrito2025understanding}. As such, the
well-known problem of hallucinations now threatens the reliability of
critical software systems~\citep{liu2023codegeneratedchatgptreally},
spurring a growing body of work on assessing the correctness of
LLM-generated code~\citep{gao2025systematicliteraturereviewcode}.
Yet, in many settings, external validation is unavailable or
incomplete: generated programs may lack reference implementations or
complete test suites~\citep{hendrycks2021apps,nguyen2022empirical,
nogueira2025beyond}. 

A common reference-free strategy is to estimate uncertainty~\citep{huang2025look} from
\emph{disagreement} across multiple independently sampled outputs. In
natural language generation, such methods interpret variation across
samples as a signal of epistemic
uncertainty~\citep{farquhar2024semantic,kuhnsemantic,
abbasi2024believe}. Recent work adapts the same principle to code
generation~\citep{li2024selective,sharma2025assessing,
valentin2026incoherence}, using signals that range from static
analysis of program text to dynamic comparison of execution
behaviour.

\paragraph{A four-axis taxonomy of sample-based uncertainty estimators.}
Although many uncertainty estimators now exist, they are
difficult to compare directly: they combine different notions of
equivalence, aggregation, reference behaviour and disagreement. Our
first contribution is therefore a taxonomy that makes these choices
explicit.

At a high level, the methods in Table~\ref{tab:prior_work_axes} share
a common sample-and-aggregate structure: they sample $K$ candidate
outputs, identify or compare behaviours using some proxy for
equivalence, and aggregate the resulting evidence into a scalar
uncertainty score. We organise this design space along four axes:
\emph{equivalence proxy}, \emph{aggregation structure},
\emph{reference behaviour} and \emph{dissimilarity}.

\emph{Equivalence proxy} specifies how candidate outputs are represented
or grouped into behaviours. Existing methods use a range of proxies,
including NLI- or LLM-based equivalence judgements
~\citep{manakul2023selfcheckgpt}, syntactic and data-flow
heuristics~\citep{li2024selective}, symbolic execution
~\citep{sharma2025assessing}, execution-based grouping on shared
inputs~\citep{valentin2026incoherence}, exact string
matching~\citep{self_cons}, and embedding similarity
~\citep{lin2023generating}.

\emph{Aggregation structure} specifies whether uncertainty is computed
from marginal behaviour probabilities or from pairwise comparisons
between behaviours. Marginal methods
~\citep{farquhar2024semantic,sharma2025assessing,self_cons} summarise
how probability mass is spread across behaviours, without asking
whether two behaviours are semantically close or far apart. Pairwise
methods~\citep{manakul2023selfcheckgpt,friel2023chainpoll,
li2024selective,valentin2026incoherence,lin2023generating}, by
contrast, explicitly compare behaviours.

\emph{Reference behaviour} applies to pairwise methods and specifies
which behaviours are compared. Some methods aggregate over all pairs
(\emph{uniform-pairwise})~\citep{manakul2023selfcheckgpt,
li2024selective,lin2023generating}; some compare the dominant
behaviour against alternatives (\emph{top-anchored})
~\citep{valentin2026incoherence}; and some compare only the
top-ranked output against alternatives while discarding alternative
probabilities (\emph{top1-only})~\citep{friel2023chainpoll}.

\emph{Dissimilarity} applies to pairwise methods and specifies how
differences between compared behaviours are scored. Existing pairwise
methods use a binary or shallow notion of disagreement: two behaviours
either agree or disagree~\citep{manakul2023selfcheckgpt,
friel2023chainpoll,valentin2026incoherence}. Marginal methods, such
as Semantic Entropy~\citep{farquhar2024semantic}, do not use an
inter-behaviour dissimilarity at all.

\begin{table}[t]
\centering
\footnotesize
\renewcommand{\arraystretch}{1.1}
\caption{Representative sample-based uncertainty estimators positioned
along four design axes. \emph{Equiv. proxy} denotes how outputs are
represented or grouped into behaviours. \emph{Aggregation} is marginal
or pairwise. \emph{Reference} specifies which behaviours are compared
for pairwise methods. \emph{Dissimilarity} specifies how compared
behaviours are scored; N/A indicates that no pairwise comparison is
used.}
\label{tab:prior_work_axes}
\begin{tabular*}{\linewidth}{@{\extracolsep{\fill}}lllll@{}}
\toprule
\textbf{Method} & \textbf{Equiv. proxy} & \textbf{Aggregation} &
\textbf{Reference} & \textbf{Dissimilarity} \\
\midrule
Semantic Entropy~\citep{farquhar2024semantic,kuhnsemantic}
  & NL-Entailment        & marginal & N/A              & N/A \\
Symbolic equiv.~\citep{sharma2025assessing}
  & symbolic execution    & marginal & N/A              & N/A \\
Self-Consistency~\citep{self_cons}
  & exact string match        & marginal & N/A     & N/A \\
SelfCheckGPT~\citep{manakul2023selfcheckgpt}
  & NL-LLM             & pairwise & uniform-pairwise & binary \\
ChainPoll~\citep{friel2023chainpoll}
  & LLM voting            & pairwise & top1-only        & binary \\

HonestCoder~\citep{li2024selective}
  & multi-modal heuristic & pairwise & uniform-pairwise  & binary \\

DiffTrust~\citep{valentin2026incoherence}
  & execution fuzz        & pairwise & top-anchored     & binary \\

EigenScore~\citep{lin2023generating}
  & NL-Embedding        & pairwise & uniform-pairwise     & binary  \\
\midrule
\rao{} (ours)
  & execution fuzz        & pairwise & uniform-pairwise & graded \\
\dcd{} (ours)
  & execution fuzz        & pairwise & top-anchored     & graded \\
\bottomrule
\end{tabular*}
\smallskip
\end{table}

\textbf{Limitation of existing methods.}
By separating design choices, the taxonomy makes prior methods easier to
compare and exposes a key limitation: \emph{existing estimators provide only
a coarse account of behavioural difference.} 
Marginal methods collapse uncertainty to the distribution of
mass over behaviours, without asking whether those behaviours are
semantically close or far apart. Pairwise methods do compare
behaviours, but typically only through a binary or shallow
agree/disagree signal. As a result,
two programs that differ on one input out of ten may be treated like
two programs that disagree on every input. This is a poor abstraction
for code, where behavioural differences are often graded
~\citep{yeo2024framework}. A useful sample-based uncertainty estimator should
therefore capture not only whether sampled programs disagree, but also
how severely they disagree.

\textbf{Our approach.}
Our key observation is that programs, unlike free-form natural
language, are executable objects. This makes semantic difference
directly measurable: two programs can be compared by how often, and in
what ways, their executions differ on shared inputs.

We therefore propose \textit{\textbf{semantic distance-aware
uncertainty estimation}}. Instead of treating all behavioural
disagreements as equally severe, we assign graded dissimilarities
between execution behaviours and aggregate them into uncertainty
scores. The framework is agnostic to the equivalence proxy: any
procedure that identifies behaviourally equivalent programs can be
used. In this paper, we use fuzzing:
unlike lightweight static
proxies~\citep{li2024selective}, it captures observable program behaviour, while avoiding the
substantial cost of more principled static techniques such as symbolic
execution~\citep{sharma2025assessing}.

We define two metrics, corresponding to two reference behaviours.
\rao{} uses uniform-pairwise aggregation, grounded in Rao's quadratic
entropy, and measures global behavioural diversity across sampled
programs. \dcd{} uses top-anchored aggregation, measuring how strongly
the alternatives disagree with the dominant behaviour containing the
top-ranked program. The former is suited to task- or model-level
uncertainty, while the latter targets the reliability of the specific
output shown to the user.

In summary, the contributions of this paper are as follows:
\begin{itemize}[noitemsep, topsep=0pt, parsep=0pt, partopsep=0pt,
leftmargin=*]

  \item We introduce a four-axis \emph{\textbf{taxonomy of sample-based uncertainty
estimators}}, making existing methods easier to compare and
clarifying which design choices distinguish them.

  \item We propose \emph{\textbf{semantic distance-aware uncertainty metrics}} as a
  proxy for correctness in code generation. On
  \emph{LiveCodeBench}, our metrics achieve \emph{\textbf{strong
  discrimination between correct and incorrect solutions}}, with
  AUROC $>0.8$.

  \item We provide extensive evidence of generalisation across
  benchmarks, languages, models, and sampling regimes. Beyond
  \emph{LiveCodeBench}, our metrics remain effective on
  \emph{\textbf{MBPP}}~\citep{MBPP_dataset},
  \emph{\textbf{BigCodeBench}}~\citep{zhuo2024bigcodebench} and
  \emph{\textbf{HumanEval-X}}~\citep{zheng2023codegeex}; across
  \emph{\textbf{Python}}, \emph{\textbf{Java}} and
  \emph{\textbf{C++}}; across closed-source and open-weights models;
  and under varying task difficulties and sampling temperatures.
  
\item We show that distance-aware uncertainty is
\emph{\textbf{cost-effective}} relative to strong state-of-the-art
uncertainty estimation baselines: our metrics substantially improve predictive
performance while reducing runtime by approximately $48$--$79\%$.

\end{itemize}

\section{Distance-Aware Uncertainty Measures}
\label{sec:metrics}

We now formalise the three components needed to compute our
distance-aware uncertainty scores: semantic clusters, distances between
clusters, and the aggregation rules that turn these distances into a
scalar uncertainty estimate. The clustering step is not specific to
our metric: any procedure that partitions candidate programs into
behaviourally equivalent groups can be used. In this paper, we use
execution signatures over a shared input set, which provides a simple
and practical instantiation.

\paragraph{Setup: Semantic Clusters.}
Given a task description $d$, let
$\Pi=\{\pi_1,\ldots,\pi_K\}$ be $K$ candidate programs sampled from a
stochastic code-generation model. Let
$\mathcal{I}=(x_1,\ldots,x_N)$ be a fixed set of inputs shared across
all candidates. Executing a program $\pi$ on $\mathcal{I}$ yields its
\emph{execution signature} $\sigma_{\mathcal{I}}(\pi) = \bigl(o_1(\pi),\ldots,o_N(\pi)\bigr)$,
where $o_k(\pi)$ is the observed outcome of $\pi$ on input $x_k$.
An outcome is either a normal output value or an abnormal termination
labelled by its error type.

Programs are placed in the same semantic cluster iff their execution
signatures are identical. Let $\{C_1,\ldots,C_M\}$ be the resulting
partition of $\Pi$. The empirical cluster probability is
  $p_i = \frac{|C_i|}{K}$, with $\sum_{i=1}^{M} p_i = 1$.
  
\subsection{Semantic Distance}
\label{sec:distance}

We next define a graded distance between semantic clusters. Since all
programs in a cluster have the same execution signature on
$\mathcal{I}$, the distance between two clusters is independent of the
choice of representatives.
For any $\pi\in C_i$ and $\pi'\in C_j$, the semantic distance between
clusters $C_i$ and $C_j$ is
\begin{equation}
  d_{ij}
  =
  \frac{1}{N}\sum_{k=1}^{N}
  \delta\bigl(o_k(\pi),o_k(\pi')\bigr).
  \label{eq:cluster-distance}
\end{equation}
The per-input outcome distance $\delta(o,o')\in[0,1]$ is defined as
\begin{equation}
  \delta(o,o')
  =
  \begin{cases}
    0 & \text{if $o$ and $o'$ are equal normal outputs,} \\
    1 & \text{if both outcomes are normal and $o\neq o'$,} \\
    a & \text{if exactly one outcome is abnormal,} \\
    b & \text{if both outcomes are abnormal with different error types,} \\
    c & \text{if both outcomes are abnormal with the same error type.}
  \end{cases}
  \label{eq:per-input-dist}
\end{equation}
where $a,b,c\in[0,1]$ are graded disagreement costs. We treat these
values as hyperparameters because abnormal executions do not induce a
canonical semantic distance: identical errors may indicate related
failure modes, but need not imply behavioural equivalence. We use a
fixed default in the main experiments (see \S\ref{sec:setup}) and assess sensitivity in
\S\ref{sec:rq4}.

By construction, $d_{ij}$ is symmetric, satisfies $d_{ii}=0$, and lies
in $[0,1]$.

\subsection{Uncertainty Metrics}
\label{sec:uncertainty-metrics}

Given semantic clusters $\{C_1,\ldots,C_M\}$, empirical probabilities
$\{p_1,\ldots,p_M\}$, and pairwise distances $\{d_{ij}\}$, we define
two distance-aware uncertainty metrics. They share the same clustering
and distance definitions, but differ in their anchor: one aggregates
globally over all cluster pairs, while the other anchors uncertainty
to the cluster containing the top-ranked output.

\paragraph{Semantic Distance Entropy (SDE).}
Our first measure is a symmetric distance-aware estimator inspired by
Rao's quadratic entropy, weighting each cluster pair by the product of
cluster probabilities and their semantic distance:
\begin{equation}
  \rao
  =
  \sum_{i < j} p_i\,p_j\,d_{ij}.
  \label{eq:sde-def}
\end{equation}
\rao{} captures global behavioural diversity: it is large when
substantial probability mass is spread across clusters that are far
apart in execution behaviour.

\paragraph{Dominant Semantic Distance Entropy (DSDE).}
In deployment, users typically inspect or rely on the model's
top-ranked output~\citep{chen2021evaluating}. We therefore define a
top-anchored variant that measures uncertainty relative to the
behaviour that would actually be served. Let
$c^*\in\{1,\ldots,M\}$ be the index of the cluster containing the
top-ranked program $\pi_1$, so that $C_{c^*}=C(\pi_1)$. We define
\begin{equation}
  \dcd
  =
  \sum_{i\neq c^*} p_i\,d_{c^*,i}.
  \label{eq:dcd-def}
\end{equation}
\dcd{} measures how strongly the alternatives disagree with the
top-ranked behaviour, weighted by the empirical probability of each
alternative. Unlike \rao{}, which treats all cluster pairs
symmetrically, \dcd{} is targeted at the reliability of the specific
output shown to the user.

\section{Overview of the Approach}
\label{sec:pipeline}

We present an end-to-end evaluation pipeline for execution-based
uncertainty estimation in code generation, which consists of four main stages: 

\textbf{First stage: Candidate program sampling.} Given a programming task described by natural language input \( d \),
let \( \text{Coder}(\cdot) \) denote a code generation system.
Without access to reference implementations or model internals,
the system independently samples \( K \) candidate programs $\{\pi_1, \pi_2, \ldots, \pi_K\}, \pi_k \sim p(\pi \mid d),$
where each $\pi_k$ corresponds to the raw code returned by API call.

\textbf{Second stage: Fuzzing-based input generation.}
Starting from an initial set of valid task inputs, we generate fuzzed
inputs using a lightweight, type-aware mutation strategy designed for
the heterogeneous input spaces common in code-generation benchmarks.
We apply structure-preserving, type-specific mutations iteratively
until \(N\) executable fuzzed inputs are obtained, providing behavioural
variation for execution-based equivalence and uncertainty estimation.

By default, we follow Valentin et al.~\citep{valentin2026incoherence}
and use benchmark-provided test cases as seed inputs. However, our
pipeline does not rely on such seeds: in \S\ref{sec:rq4}, we show that
a seed-free variant remains effective. 
We also report input-quality
diagnostics for the resulting fuzz suites, including validity,
uniqueness, and coverage, in Appendix~\ref{app:input-quality-detail}.

More generally, the pipeline is
\emph{fuzzer-agnostic}: the input-generation component can be replaced
by any suitable fuzzing strategy, allowing us to inherit advances from
the fuzzing community. For example, in \S\ref{sec:rq2}, we use
whitebox fuzzing~\citep{godefroid2008automated} for BigCodeBench.

\textbf{Third stage: Execution-based representation and clustering.}
Let $\mathbf{i} = (i_1, i_2, \ldots, i_N)$ denote the set of test inputs associated
with the task, where $N$ is the number of inputs.
This input set is shared across all candidate programs
$\boldsymbol{\pi} = (\pi_1, \pi_2, \ldots, \pi_K)$.
Each candidate program $\pi_k$ is executed on all inputs in $\mathbf{i}$, yielding
an execution output vector $\mathbf{o}_k = \bigl( \pi_k(i_1), \pi_k(i_2), \ldots, \pi_k(i_N) \bigr).$

Candidate programs are clustered according to behavioural equivalence: programs
with identical execution signatures are assigned to the same behavioural cluster.
For abnormal executions, programs are grouped into the same cluster if they
terminate with the same error.

\textbf{Fourth stage: Uncertainty computation.}
Based on the semantic clusters, we compute \rao{} and \dcd{}. 

\section{Experiments}
\label{sec:experiments}

In this section, we will evaluate our \raodcd{}. 
Our experiments are designed to answer the following questions: 

\textbf{RQ1 (Predictive Performance).}
Do \raodcd{} reliably estimate program correctness, and how do they
compare to existing 
baselines~\citep{valentin2026incoherence,li2024selective,farquhar2024semantic}?

\textbf{RQ2 (Generalisation).}
Do \raodcd{} transfer across benchmarks of varying difficulty and
scale, programming languages, sampling temperatures and model
families?

\textbf{RQ3 (Cost-effectiveness).}
Are \raodcd{} cheap enough for inference-time deployment under
aggressive sample budgets and competitive against runtime baselines?

\textbf{RQ4 (Robustness to design choices).}
Do \raodcd{} remain reliable under different design choices, including
fixed versus tuned distance weights and the absence of
benchmark-provided test inputs?

\subsection{Experimental Setup}
\label{sec:setup}

\textbf{Models.}
We conduct experiments using representative code generation models
spanning different capability tiers: GPT-3.5-Turbo (low), GPT-4o-mini
(medium), Gemini-2.5-Flash-Lite (medium), and Claude Opus~4.5 (high),
with respective \emph{pass@1} success rates of 30.83\%, 49.24\%, 60.53\%,
and 82.73\% on \emph{LiveCodeBench}. To assess generalisation
(\S\ref{sec:rq2}), we additionally use the open-weights model
DeepSeek-Coder-V2 (\emph{pass@1} 38.14\% on \emph{LiveCodeBench}).

\textbf{Datasets.} 
We evaluate on four complementary benchmarks spanning different
task difficulties and programming languages.
Our primary benchmark is \emph{LiveCodeBench}
(LCB)~\citep{livecodebench} (Python). To assess generalisation
(\S\ref{sec:rq2}), we additionally use \emph{MBPP}~\citep{MBPP_dataset}
(Python), \emph{HumanEval-X}~\citep{zheng2023codegeex} (Python, Java,
and C++), and \emph{BigCodeBench}~\citep{zhuo2024bigcodebench}
(Python). Detailed statistics and task descriptions are in
Appendix~\ref{app:datasets}. 

\textbf{Experimental configuration.}
Unless otherwise specified, we sample with temperature $T = 0.6$
following prior code-generation work~\citep{li2024selective,
guo2025deepseek}, generate $K = 10$ candidate programs per task, and
evaluate them on $N = 10$ fuzz inputs with a $0.2$\,s per-input
execution timeout. We use $(a, b, c) = (1, 0.8, 0.6)$ from \S\ref{sec:distance}; a sensitivity analysis shows the fixed weights deviate by less than $0.011$ AUROC from per-setting learned weights, more details can be found in \S\ref{sec:rq3}.
All experiments are run on an Apple M4 Pro with
24\,GB of RAM.

\textbf{Prediction targets.}
Following prior work~\citep{chen2021evaluating}, we adopt \emph{pass@1} as a primary
correctness signal, capturing whether the model succeeds on its first attempt.
Since \emph{pass@1} is binary, it does not distinguish near-correct programs from
substantially incorrect ones. We therefore use \emph{\tpass} as a
complementary, graded execution-based target, defined as the proportion of
test cases passed by the first sampled program, as in~\citep{sharma2025assessing}.
This provides a useful measure of \emph{partial correctness}.

\textbf{Evaluation metrics.}
For \emph{pass@1}, which is a binary outcome, we evaluate how well uncertainty scores
\emph{discriminate} between correct and incorrect generations using the area
under the ROC curve (AUROC)~\citep{bradley1997use}.
For the continuous-valued \emph{\tpass}{},
we assess its association with uncertainty scores using Pearson \citep{benesty2009pearson}
 and Spearman \citep{wissler1905spearman} correlation coefficients, capturing linear and monotonic relationships, respectively.

\subsection{Answering RQ1 (Predictive Performance)}
\label{sec:rq1}

\textbf{\emph{pass@1} and \emph{\tpass}{} prediction.}
Table~\ref{tab:main_results} reports AUROC for \emph{pass@1} failure
prediction and Pearson $r$/Spearman $\rho$ correlations with \emph{\tpass}{}
on \emph{LiveCodeBench}. Across all four models, \rao{} and
\dcd{} provide strong predictive signals: \dcd{} achieves AUROC above
$0.8$ in every setting, and both metrics correlate negatively with
\emph{\tpass}{}, indicating that higher uncertainty corresponds to lower
partial correctness.

On the main closed-source LiveCodeBench evaluation,
\dcd{} improves over \rao{} across all models and metrics, supporting
the top-anchored design: when assessing the program
shown to the user, uncertainty can be more informative when measured
relative to that program's behaviour rather than averaged uniformly
across all samples.  This advantage may depend on
the first sample being representative; when it is not, as suggested by
DeepSeek-Coder-V2 in Table~\ref{tab:generalisation}, the symmetric
\rao{} estimator can be more stable.

\textbf{Comparison with baselines.}
We compare against three strong baselines, chosen to cover
the main existing design choices in Table~\ref{tab:prior_work_axes}:
DiffTrust, the state-of-the-art incoherence-based proxy~\citep{valentin2026incoherence};
HonestCoder, a representative confidence estimator combining embeddings,
data-flow signatures, and syntactic $n$-grams~\citep{li2024selective};
and Semantic Entropy, the standard entropy-based uncertainty estimator
over LLM-judged semantic clusters~\citep{farquhar2024semantic}. As
shown in Table~\ref{tab:main_results}, even these competitive baselines
are only weakly informative on \emph{LiveCodeBench}. By contrast,
\rao{} and \dcd{} outperform every baseline on every model and metric,
suggesting that graded semantic distance captures correctness-relevant
information missed by binary disagreement and marginal entropy-based
uncertainty.

\begin{table*}[h]
\centering
\footnotesize
\setlength{\tabcolsep}{3pt}
\renewcommand{\arraystretch}{1.1}
\caption{Performance of \rao{}, \dcd{}, and the three 
baselines on \emph{LiveCodeBench} across models ($T=0.6$). Best
result per metric per row is shown in bold. AUC denotes AUROC; $r$ and
$\rho$ are Pearson and Spearman correlations with \emph{\tpass}{}, respectively.
Per-difficulty breakdowns (Easy/Medium/Hard) are reported in
Appendix~\ref{app:per-difficulty-tables}. }
\label{tab:main_results}
\resizebox{\textwidth}{!}{%
\begin{tabular}{l ccc ccc | ccc ccc ccc}
\toprule
& \multicolumn{3}{c}{\textbf{\rao{}}}
& \multicolumn{3}{c}{\textbf{\dcd{}}}
& \multicolumn{3}{c}{\textbf{DiffTrust}}
& \multicolumn{3}{c}{\textbf{HC}}
& \multicolumn{3}{c}{\textbf{Semantic Entropy}} \\
\cmidrule(lr){2-4} \cmidrule(lr){5-7} \cmidrule(lr){8-10} \cmidrule(lr){11-13} \cmidrule(lr){14-16}
\textbf{Model}
& AUC & $r$ & $\rho$
& AUC & $r$ & $\rho$
& AUC & $r$ & $\rho$
& AUC & $r$ & $\rho$
& AUC & $r$ & $\rho$ \\
\midrule
GPT-3.5-Turbo-0125
  & 0.827 & $-$0.587 & $-$0.581
  & \textbf{0.844} & \textbf{$-$0.622} & \textbf{$-$0.620}
  & 0.669 & $-$0.259 & $-$0.267
  & 0.646 & $-$0.344 & $-$0.328
  & 0.605 & $-$0.240 & $-$0.237 \\
GPT-4o-mini
  & 0.826 & $-$0.589 & $-$0.584
  & \textbf{0.844} & \textbf{$-$0.624} & \textbf{$-$0.624}
  & 0.534 & $-$0.024 & $-$0.064
  & 0.646 & $-$0.344 & $-$0.328
  & 0.607 & $-$0.240 & $-$0.238 \\
Gemini-2.5-Flash-Lite
  & 0.794 & $-$0.494 & $-$0.491
  & \textbf{0.808} & \textbf{$-$0.532} & \textbf{$-$0.512}
  & 0.688 & $-$0.296 & $-$0.298
  & 0.668 & $-$0.301 & $-$0.329
  & 0.639 & $-$0.237 & $-$0.242 \\
Claude-opus-4.5
  & 0.818 & $-$0.552 & $-$0.450
  & \textbf{0.825} & \textbf{$-$0.611} & \textbf{$-$0.461}
  & 0.606 & $-$0.139 & $-$0.139
  & 0.739 & $-$0.424 & $-$0.350
  & 0.687 & $-$0.275 & $-$0.287 \\
\bottomrule
\end{tabular}%
}
\end{table*}

\subsection{Answering RQ2 (Generalisation)}
\label{sec:rq2}

\textbf{Across benchmarks of varying difficulty.}
The first block of Table~\ref{tab:generalisation} evaluates \rao{}
and \dcd{} on three Python benchmarks spanning entry-level function
completion (MBPP), competitive programming (\emph{LiveCodeBench}), and
repository-scale tasks (BigCodeBench). Both metrics retain strong
predictive performance on MBPP and \emph{LiveCodeBench}, with AUROC above
$0.75$ and Spearman correlations close to $-0.6$.

\textit{Beyond function-level benchmarks: BigCodeBench.}
BigCodeBench is the most demanding benchmark we consider and, to our
knowledge, the first repository-scale evaluation of
sample-based uncertainty estimation for LLM-generated code. 
Its tasks include cross-file
dependencies, side effects, stochastic operations, and often a range
of valid outputs for the same specification. This makes exact-match
execution clustering too brittle: semantically valid programs may
produce different concrete outputs and be incorrectly separated.

Our formulation can accommodate this setting because it is agnostic with respect to
the clustering proxy. We therefore make
a single drop-in change: instead of clustering by exact output match,
we use whitebox fuzzing~\citep{godefroid2008automated} to target
specification predicates and group programs by predicate satisfaction.
With this adaptation, \dcd{} achieves AUROC $0.668$ and Spearman
$\rho=-0.335$ on BigCodeBench. Although lower than on function-level
benchmarks, this remains a non-trivial signal in a substantially more
challenging repository-scale regime.

\textbf{Across programming languages.}
On HumanEval-X~\citep{zheng2023codegeex}, covering Python, Java, and
C++, both metrics perform consistently across languages without
language-specific tuning.

\textbf{Across sampling temperatures.}
We sweep the sampling temperature $T \in \{0.2,0.6,0.8\}$ using
GPT-4o-mini on \emph{LiveCodeBench}. As shown in Table~\ref{tab:generalisation},
performance improves with temperature: lower temperatures reduce
sample diversity, while higher temperatures expose richer behavioural
variation for uncertainty estimation. Nevertheless, \rao{} and \dcd{}
remain informative across all settings, indicating that distance-aware
uncertainty benefits from, but does not depend on, high sampling
diversity.

\textbf{Across model families.}
Finally, we evaluate on the open-weights DeepSeek-Coder-V2 model,
which achieves \emph{pass@1} of $38.14\%$ on \emph{LiveCodeBench}. Both metrics
retain meaningful AUROC and correlations with \emph{\tpass}{}, showing that
the distance-aware signal transfers beyond the closed-source models
used in RQ1.
\begin{table}[h]
\centering
\scriptsize
\setlength{\tabcolsep}{3pt}
\renewcommand{\arraystretch}{1.0}
\caption{Performance of \rao{} and \dcd{} across benchmarks,
languages, sampling temperatures, and model families
(\texttt{GPT-4o-mini}, $T=0.6$, unless noted). Better of
\rao{}/\dcd{} per row in AUROC in bold. Per-difficulty
\emph{LiveCodeBench} breakdowns are in
Appendix~\ref{app:per-difficulty-tables}.}
\label{tab:generalisation}
\begin{tabularx}{\linewidth}{l*{6}{>{\centering\arraybackslash}X}}
\toprule
& \multicolumn{2}{c}{AUROC}
& \multicolumn{2}{c}{Pearson $r$}
& \multicolumn{2}{c}{Spearman $\rho$} \\
\cmidrule(lr){2-3} \cmidrule(lr){4-5} \cmidrule(lr){6-7}
\textbf{Benchmark}
& \rao{} & \dcd{} & \rao{} & \dcd{} & \rao{} & \dcd{} \\
\midrule
\multicolumn{7}{l}{\textit{Increased difficulty (Python benchmarks, easy to hard)}} \\
\midrule
MBPP              & 0.751 & \textbf{0.752} & $-$0.480 & $-$0.434 & $-$0.601 & $-$0.595 \\
LiveCodeBench     & 0.826 & \textbf{0.844} & $-$0.589 & $-$0.624 & $-$0.584 & $-$0.624 \\
BigCodeBench      & 0.664 & \textbf{0.668} & $-$0.281 & $-$0.287 & $-$0.333 & $-$0.335 \\
\midrule
\multicolumn{7}{l}{\textit{Programming languages (HumanEval-X)}} \\
\midrule
Python            & 0.751 & \textbf{0.757} & $-$0.523 & $-$0.521 & $-$0.631 & $-$0.634 \\
Java              & 0.740 & \textbf{0.745} & $-$0.509 & $-$0.516 & $-$0.605 & $-$0.603 \\
C++               & 0.796 & \textbf{0.804} & $-$0.599 & $-$0.619 & $-$0.573 & $-$0.587 \\
\midrule
\multicolumn{7}{l}{\textit{Sampling temperature (GPT-4o-mini, LCB)}} \\
\midrule
$T = 0.2$         & \textbf{0.780} & 0.779 & $-$0.538 & $-$0.529 & $-$0.524 & $-$0.522 \\
$T = 0.6$         & 0.826 & \textbf{0.844} & $-$0.589 & $-$0.624 & $-$0.584 & $-$0.624 \\
$T = 0.8$         & 0.837 & \textbf{0.853} & $-$0.612 & $-$0.660 & $-$0.609 & $-$0.646 \\
\midrule
\multicolumn{7}{l}{\textit{Open-weights model (LCB)}} \\
\midrule
DeepSeek-Coder-V2 & \textbf{0.778} & 0.769 & $-$0.483 & $-$0.469 & $-$0.479 & $-$0.462 \\
\bottomrule
\end{tabularx}
\end{table}

\subsection{Answering RQ3 (Cost-effectiveness)}
\label{sec:rq3}

\begin{figure}[h]
\centering
\includegraphics[width=\linewidth]{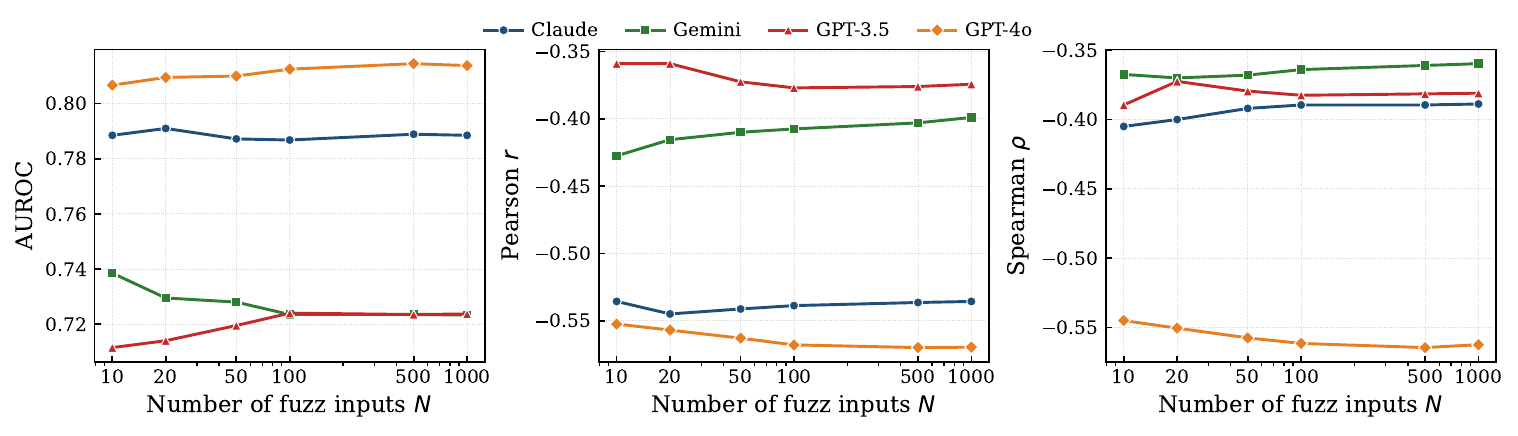}
\caption{Sensitivity of \dcd{} to the number of fuzz inputs $N$
on a 50-task \emph{LiveCodeBench} subset, with $K$ fixed to $10$. AUROC,
Pearson $r$, and Spearman $\rho$ remain essentially flat across
$N \in \{10, 20, 50, 100, 500, 1000\}$ for all four
code-generation models.}
\label{fig:sensitivity_N}
\end{figure}

\textbf{Scaling $K$ and $N$ up to test for further gains.}
To check whether allocating a larger budget improves predictive
performance, we extend the sensitivity analysis on a 50-task
\emph{LiveCodeBench} subset across all four code-generation models.
Sweeping $K \in \{5, 10, 15, 20\}$ at fixed $N = 10$ leaves
\dcd{} AUROC within $0.024$ of the default $(K, N) = (10, 10)$ across all values;
sweeping $N$ from $10$ up to $1{,}000$ at fixed $K = 10$ likewise
produces no measurable trend (Figure~\ref{fig:sensitivity_N}).
Both sweeps indicate that scaling $K$ or $N$ beyond the default
$(K, N) = (10, 10)$ yields no measurable predictive gain.

\begin{figure}[h]
\centering
\includegraphics[width=\linewidth]{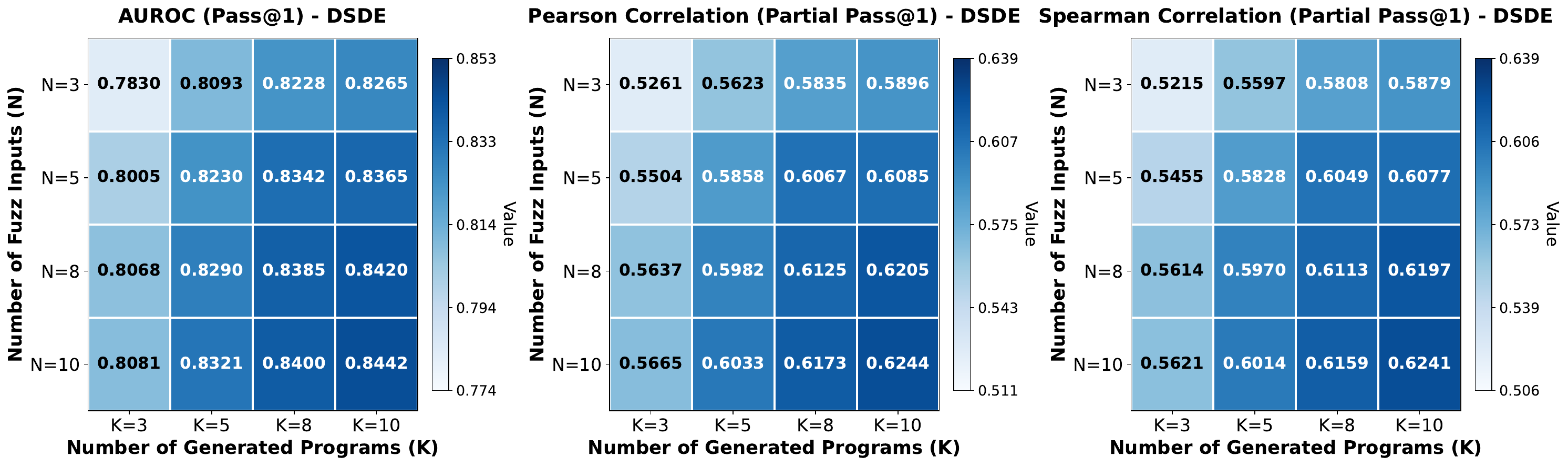}
\caption{Joint cost-effectiveness analysis of \dcd{} on \emph{LiveCodeBench}
(GPT-4o-mini, $T = 0.6$), varying $K \in \{3, 5, 8, 10\}$ and
$N \in \{3, 5, 8, 10\}$.}
\label{fig:KN_heatmap}
\end{figure}

\textbf{Scaling $K$ and $N$ down to find the minimal viable configuration.}
Since enlarging the budget does not help, we instead probe how far
$(K, N)$ can be reduced before the predictive signal degrades. We
sweep $K, N \in \{3, 5, 8, 10\}$ jointly
(Figure~\ref{fig:KN_heatmap}). The smallest configuration $(K, N)
= (3, 3)$ already yields AUROC $0.783$, recovering ${\sim}93\%$ of
the AUROC at $(10, 10)$ while requiring an order-of-magnitude less
sample and execution cost. Therefore,  $(3, 3)$ can be adopted as a
cost-effective deployment default for budget-constrained settings.

\textbf{Runtime against baselines.}
We now compare
end-to-end wall-clock cost against other uncertainty
estimators. Table~\ref{tab:costtime_breakdown} breaks runtime into:
candidate-program sampling, additional model-side computation
(\emph{Model Cost}, e.g. embeddings or LLM-as-judge calls), fuzz-input
generation, program execution and metric computation. The sampling
cost is shared across methods; the main differences arise from the
additional model-side or execution costs required by each estimator.

At the default $(K,N)=(10,10)$, our method takes on average ${\sim}5.7$\,s per
task, compared with ${\sim}11$\,s for HonestCoder, ${\sim}13$\,s for
Semantic Entropy, and ${\sim}27$\,s for DiffTrust. This is a
${\sim}79\%$ reduction relative to state-of-the-art DiffTrust. With the deployment
default $(K,N)=(3,3)$, the runtime falls further to ${\sim}2$\,s per
task.

\begin{table}[h]
\centering
\scriptsize
\setlength{\tabcolsep}{3pt}
\renewcommand{\arraystretch}{1.0}
\caption{Per-task runtime breakdown on \emph{LiveCodeBench} with \texttt{GPT-4o-mini}.}
\label{tab:costtime_breakdown}
\begin{tabular*}{\linewidth}{@{\extracolsep{\fill}}lccccccc@{}}
\toprule
\textbf{Method} & $\boldsymbol{N}$
  & \textbf{Sampling} & \textbf{Model Cost}
  & \textbf{Fuzzing} & \textbf{Execution}
  & \textbf{Metric} & \textbf{Total} \\
\midrule
HonestCoder   & ---  & ${\sim}3$\,s & ${\sim}7$\,s  & ---          & ---            & ${\sim}1$\,s & ${\sim}11$\,s \\
Semantic Entropy         & ---  & ${\sim}3$\,s & ${\sim}10$\,s & ---          & ---            & ${<}1$\,ms   & ${\sim}13$\,s \\
DiffTrust     & 1000 & ${\sim}3$\,s & ---           & ${\sim}3$\,s & ${\sim}20$\,s  & ${\sim}1$\,s & ${\sim}27$\,s \\
\midrule
\textbf{Ours} & 10   & ${\sim}3$\,s & ---           & ${<}5$\,ms   & ${\sim}2.7$\,s & ${<}1$\,ms   & $\boldsymbol{{\sim}5.7}$\,\textbf{s} \\
\bottomrule
\end{tabular*}
\end{table}

The saving comes from two design choices: (i) \raodcd{} avoid
model costs entirely (no embeddings, no LLM-as-judge); and
(ii) while DiffTrust requires $N = 1{,}000$ test inputs by default
to satisfy the variance bounds of its Monte Carlo incoherence
estimator, our metric is a deterministic pairwise aggregate over
execution clusters and does not depend on a large $N$ for
statistical guarantees.

\subsection{Answering RQ4 (Robustness to Design Choices)}
\label{sec:rq4}

\textbf{Robustness to graded disagreement costs $(a,b,c)$.}
The graded disagreement costs $(a,b,c)$ are fixed to $(1, 0.8, 0.6)$
throughout the experiments above. To quantify the gap between this
fixed scheme and a per-setting tuned alternative, we additionally fit
$(a,b,c)$ via an 80:20 train--test split per model on \emph{LiveCodeBench}
at $T = 0.6$. As shown in Table~\ref{tab:robustness_weights}, the
default and learned schemes differ by less than $0.011$ AUROC across
all four models, indicating that the fixed weights are sufficient on
\emph{LiveCodeBench}. In our current experiments, the default setting appears to be stable and strong, but learned weights may still be useful in future settings or datasets with different failure patterns.

\begin{table*}[h]
\centering
\scriptsize
\setlength{\tabcolsep}{2pt}
\renewcommand{\arraystretch}{0.95}
\caption{Robustness to distance weights $(a,b,c)$: fixed default
$(1.0, 0.8, 0.6)$ vs.\ per-setting learned weights on
\emph{LiveCodeBench} (\texttt{GPT-4o-mini}, $T = 0.6$, 80:20 train--test
split per model).}
\label{tab:robustness_weights}
\begin{tabular*}{\linewidth}{@{\extracolsep{\fill}}lcccc@{}}
\toprule
Setting & $(a, b, c)$ & Default & Learned & $\Delta$ \\
\midrule
Claude Opus 4.5        & $(0.90, 0.90, 0.25)$ & 0.8248 & 0.8206 & $-$0.0042 \\
Gemini-2.5-Flash-Lite  & $(0.90, 0.90, 0.90)$ & 0.8075 & 0.7974 & $-$0.0101 \\
GPT-3.5-Turbo          & $(0.97, 0.88, 0.88)$ & 0.8442 & 0.8395 & $-$0.0047 \\
GPT-4o-mini            & $(1.00, 0.80, 0.80)$ & 0.8442 & 0.8524 & $+$0.0082 \\
\bottomrule
\end{tabular*}
\end{table*}

\textbf{Robustness under seed-free input generation
(no dataset-provided test inputs).}
In contrast to the seeded experiments above (where the fuzzer
mutates benchmark-provided test inputs as initial seeds), we further
evaluate a \emph{seed-free} variant in which the fuzzer never reads
benchmark-provided test cases, expected outputs, or reference
implementations. Parameter types are inferred by a
rule-based parser from the function interface, using type
annotations when available and parameter names or natural-language
hints from the prompt otherwise. Inputs are then sampled from
generic type-level distributions. As shown in
Table~\ref{tab:seedless}, the AUROC degradation is small (at most
$0.087$) and the predictive signal remains substantially above
chance across all benchmarks, confirming that \rao{} and \dcd{} are
fuzzer-agnostic.

\begin{table}[h]
\centering
\scriptsize
\setlength{\tabcolsep}{3pt}
\renewcommand{\arraystretch}{1.0}
\caption{Robustness to fuzz-input seeds: AUROC, Pearson, and
Spearman across three fuzzer seeds on \emph{LiveCodeBench}
(\texttt{GPT-4o-mini}, $T = 0.6$).}
\label{tab:seedless}
\begin{tabular*}{\linewidth}{@{\extracolsep{\fill}}lccc|ccc@{}}
\toprule
 & \multicolumn{3}{c|}{\textbf{\rao{}}} & \multicolumn{3}{c}{\textbf{\dcd{}}} \\
\cmidrule(lr){2-4} \cmidrule(lr){5-7}
\textbf{Benchmark}
 & Seed & Seed-free & $\Delta$
 & Seed & Seed-free & $\Delta$ \\
\midrule
MBPP                 & 0.751 & 0.712 & $-0.039$ & 0.752 & 0.711 & $-0.041$ \\
LiveCodeBench        & 0.826 & 0.786 & $-0.040$ & 0.844 & 0.787 & $-0.057$ \\
HumanEval-X (Python) & 0.751 & 0.705 & $-0.046$ & 0.757 & 0.704 & $-0.053$ \\
HumanEval-X (Java)   & 0.748 & 0.725 & $-0.023$ & 0.745 & 0.731 & $-0.014$ \\
HumanEval-X (C++)    & 0.796 & 0.709 & $-0.087$ & 0.804 & 0.723 & $-0.081$ \\
\bottomrule
\end{tabular*}
\end{table}

\textbf{Robustness to fuzz-input random seeds.}
Here, ``seed'' refers to the random seed of the fuzz-input generator,
not to the benchmark-provided seed inputs used to initialise fuzzing.
We keep the tasks, prompts, sampled programs, and benchmark-provided
initial inputs fixed, and rerun only the fuzzing step with three
different random seeds. This isolates the effect of randomness in the
mutation process itself.
We re-run the \emph{LiveCodeBench} pipeline on
\texttt{GPT-4o-mini} ($T = 0.6$, $K = N = 10$) under three
independent fuzz-input-generator seeds. As shown in
Table~\ref{tab:robustness_seed_reliability}, cross-seed standard deviations
are at most $0.0024$ on AUROC and $0.0050$ on Spearman for both
\rao{} and \dcd{}, indicating that the reported numbers do not
depend on the choice of seed.

\begin{table}[h]
\centering
\scriptsize
\setlength{\tabcolsep}{4pt}
\renewcommand{\arraystretch}{1.0}
\caption{Robustness to fuzz-input-generator randomness on
\emph{LiveCodeBench} (\texttt{GPT-4o-mini}, $T=0.6$, $K=N=10$).
The random seed controlling fuzz-input mutations is
varied. Results report mean $\pm$ standard deviation over three
independent fuzz-input-generator seeds.}
\label{tab:robustness_seed_reliability}
\begin{tabular}{lccc}
\toprule
\textbf{Metric} & \textbf{AUROC} & \textbf{Pearson} & \textbf{Spearman} \\
\midrule
\rao{}   & $0.8238 \pm 0.0011$         & $-0.5880 \pm 0.0035$         & $-0.5831 \pm 0.0022$ \\
\dcd{}  & $\mathbf{0.8413 \pm 0.0005}$ & $\mathbf{-0.6249 \pm 0.0030}$ & $\mathbf{-0.6240 \pm 0.0023}$ \\
\bottomrule
\end{tabular}
\end{table}
\section{Limitations}
\label{sec:limitations}

Our framework inherits several limitations from its underlying
assumptions. (i) \emph{Equivalence proxy.} Effectiveness depends on the proxy
exposing meaningful behavioural differences; weak fuzz inputs may yield
overly coarse equivalence classes.
(ii) \emph{Fuzzing is not a correctness oracle.} Two incorrect
programs may agree on all sampled inputs but diverge elsewhere, and
two valid programs may disagree when the specification admits
multiple acceptable behaviours, as seen on BigCodeBench.
(iii) \emph{Top-ranked reference.} When the top-ranked sample is
unrepresentative of the model's behavioural distribution, anchoring
becomes unstable, consistent with our DeepSeek-Coder-V2 result
where \rao{} slightly outperforms \dcd{}.
(iv) \emph{Heuristic graded distance.} The abnormal-execution costs
$(a, b, c)$ are hyperparameters with no canonical value across
failure modes. Weights may need adaptation for other runtimes,
languages, or benchmarks.
\section{Conclusions}
\label{sec:conclusions}

We introduced a four-axis taxonomy of sample-based uncertainty
estimators (Table~\ref{tab:prior_work_axes}) and identified an
unoccupied region: pairwise uncertainty estimation with graded
semantic dissimilarity. We instantiated this region with two estimators:
\rao{}, a symmetric Rao quadratic entropy over execution clusters,
and \dcd{}, a top-anchored variant measuring disagreement between
alternative samples and the served output. Both improve predictive
performance over sample-based baselines at lower computational cost,
and remain stable across benchmarks, languages, temperatures and
fuzzing settings.


\bibliographystyle{plainnat}
\bibliography{ref}

\newpage
\appendix
\section{Detailed Input-Quality Diagnostics}
\label{app:input-quality-detail}

The reliability of \rao{} and \dcd{} ultimately depends on whether
the generated test inputs $\mathcal{I}$ produce useful behavioural
signal on the sampled programs. This appendix reports four
diagnostics that characterise this signal, computed per task over
$\mathcal{I} = \{x_1, \dots, x_{10}\}$ and the candidate set
$\mathcal{C}$, and averaged across tasks. We organise the
diagnostics into two groups: (i) input-side validity metrics that
measure how well $\mathcal{I}$ is constructed; (ii) coverage
metrics that measure how thoroughly $\mathcal{I}$ exercises the
candidate code.

\subsubsection*{Input validity metrics}

\textbf{ValidExecRate} is the fraction of inputs on which at least
one candidate program returns a non-crash output. An input on
which every candidate crashes carries no behavioural signal and
cannot contribute to clustering:
\begin{equation}
\mathrm{ValidExecRate}
\;=\;
\frac{|\{\,x_j \in \mathcal{I} :
        \exists\, c \in \mathcal{C},\;
        c(x_j)\text{ is non-crash}\,\}|}
     {|\mathcal{I}|}.
\end{equation}

\textbf{UniqueInputRate} is the fraction of inputs that remain
distinct after canonical normalisation. A value below one
indicates that the fuzzer has produced redundant or equivalent
inputs within the same task:
\begin{equation}
\mathrm{UniqueInputRate}
\;=\;
\frac{\bigl|\,\mathrm{unique}\!\left(\{\mathrm{normalize}(x_j) : x_j \in \mathcal{I}\}\right)\bigr|}
     {|\mathcal{I}|}.
\end{equation}

\textbf{CrashPollutionRate} is the fraction of all
candidate-by-input execution cells that crash. A high rate
indicates that many cells contribute no meaningful behavioural
signal and therefore add noise to clustering:
\begin{equation}
\mathrm{CrashPollutionRate}
\;=\;
\frac{|\{\,(c, x_j) \in \mathcal{C} \times \mathcal{I} :
        c(x_j)\text{ crashes}\,\}|}
     {|\mathcal{C}|\cdot|\mathcal{I}|}.
\end{equation}

\subsubsection*{Code-coverage metrics}

\textbf{LineCoverage} and \textbf{BranchCoverage} measure the
fraction of executable lines and branches in the candidate
programs that are exercised by $\mathcal{I}$. For each candidate
$c \in \mathcal{C}$, let $L(c)$ and $L_{\mathrm{cov}}(c)$ denote
the total and covered executable lines when $c$ is executed on
$\mathcal{I}$, and let $B(c)$ and $B_{\mathrm{cov}}(c)$ denote the
analogous quantities for branches. The micro-averaged coverages
are
\begin{equation}
\mathrm{LineCoverage}
\;=\;
\frac{\sum_{c\in\mathcal{C}} L_{\mathrm{cov}}(c)}
     {\sum_{c\in\mathcal{C}} L(c)},
\qquad
\mathrm{BranchCoverage}
\;=\;
\frac{\sum_{c\in\mathcal{C}} B_{\mathrm{cov}}(c)}
     {\sum_{c\in\mathcal{C}} B(c)}.
\end{equation}
Together they bound how much of each candidate's behaviour is
observable through the fuzz suite.

\begin{table}[h]
\centering
\small
\setlength{\tabcolsep}{4pt}
\renewcommand{\arraystretch}{1.1}
\caption{Input-quality diagnostics on \emph{LiveCodeBench} at $T=0.6$.
Per task we use the first $N=10$ available inputs; tasks with no
available inputs are excluded. Each metric is averaged over the
remaining tasks. LineCov and BranchCov are micro-averaged.}
\label{tab:input_quality}
\begin{tabular}{lccccc}
\toprule
\textbf{Model} & \textbf{ValidExec} & \textbf{Unique}
& \textbf{Crash}
& \textbf{LineCov} & \textbf{BranchCov} \\
\midrule
GPT-3.5-Turbo          & 0.901 & 0.948 & 0.209 & 0.937 & 0.888 \\
GPT-4o-mini            & 0.898 & 0.948 & 0.213 & 0.937 & 0.889 \\
Gemini-2.5-Flash-Lite  & 0.851 & 0.951 & 0.269 & 0.919 & 0.874 \\
Claude-opus-4.5        & 0.863 & 0.950 & 0.201 & 0.917 & 0.879 \\
\bottomrule
\end{tabular}
\end{table}

\subsubsection*{Results.}
Table~\ref{tab:input_quality} reports the diagnostics on
\emph{LiveCodeBench} at $T=0.6$. UniqueInputRate is approximately $0.95$
across all four models, and ValidExecRate stays above $0.85$,
indicating that the fuzzer rarely produces redundant inputs and
that most inputs successfully exercise at least one candidate
program. LineCoverage exceeds $0.91$ and BranchCoverage exceeds
$0.87$ on every model, indicating that the fuzz suite reaches
most code paths in each candidate. The input pool therefore
remains diverse, mostly executable, and largely covering across
all four models. CrashPollutionRate sits around $0.20$--$0.27$,
indicating that roughly one in five execution cells contributes
no behavioural signal; reducing this rate (e.g.\ via type-aware
input constraints or LLM-guided fuzzers) is a direct lever for
strengthening the upstream signal that \rao{} and \dcd{} consume.

\section{End-to-End Evaluation Pipeline}
\label{app:pipeline}

\begin{figure*}[h]
    \centering
    \includegraphics[width=0.9\linewidth]{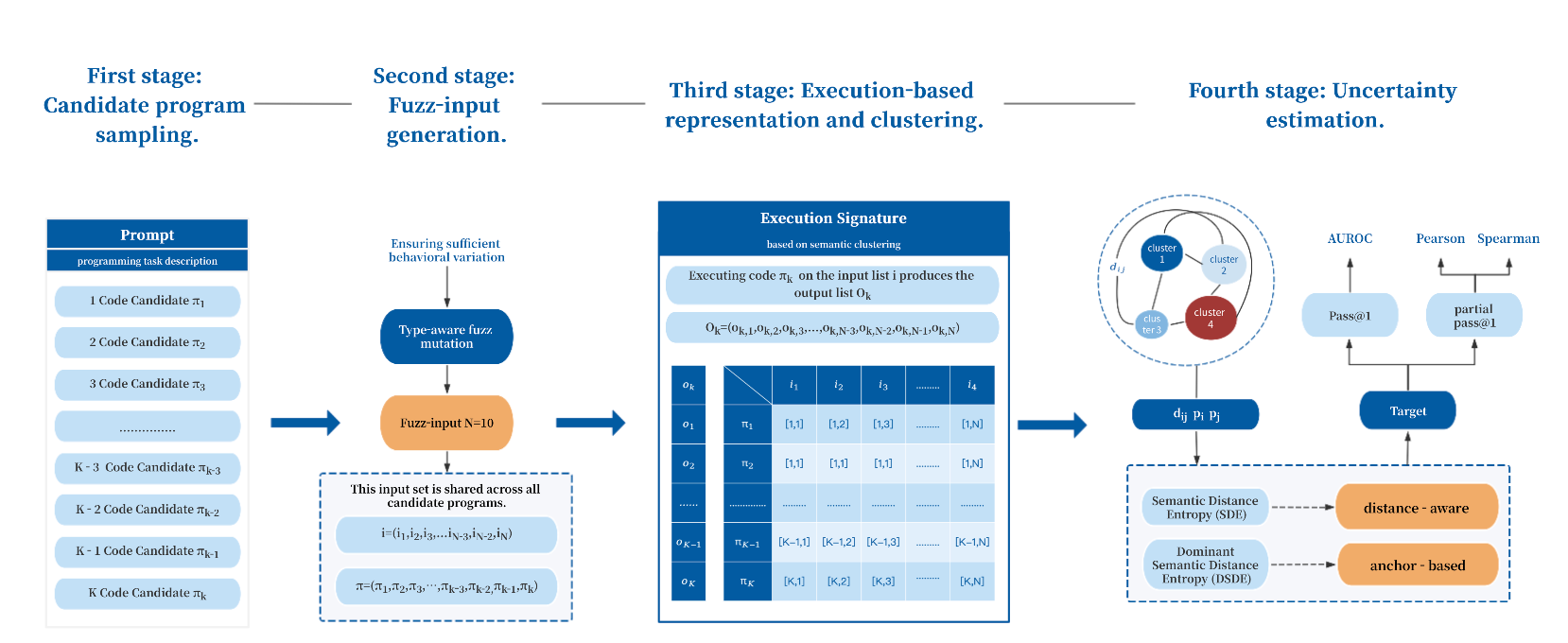}
    \caption{End-to-end evaluation pipeline for execution-based uncertainty
    estimation in code generation.}
    \label{fig:lcb-uncertainty-pipeline}
\end{figure*}

Figure~\ref{fig:lcb-uncertainty-pipeline} summarises the end-to-end
evaluation pipeline used in this paper. Candidate programs are first sampled
from the code generation model; type-aware mutations then produce a shared
set of inputs; each candidate is executed on this input set to obtain
an execution signature; programs with identical signatures are clustered; and
finally the distance-aware uncertainty metrics \rao{} and \dcd{} are computed
from the resulting cluster distribution.

\section{Extended Experimental Results}
\label{app:extended}

This appendix presents additional experimental analyses that complement the
main evaluation in Section~\ref{sec:experiments}. Unless stated otherwise,
all settings follow the protocol used in the main paper: $K=10$ sampled
programs and $N=10$ test inputs per task, with execution timeout $0.2$\,s
and sampling temperature $T=0.6$.

\subsection{Extended Per-Difficulty Tables}
\label{app:per-difficulty-tables}

Table~\ref{tab:auroc_performance_by_difficulty} and Table~\ref{tab:correlation_by_difficulty}  correspond directly
to the AUROC and correlation tables in the main paper, evaluated on four
closed-source code-generation models — \texttt{GPT-3.5-Turbo-0125},
\texttt{GPT-4o-mini}, \texttt{Gemini-2.5-Flash-Lite}, and
\texttt{Claude-opus-4.5} — with results for an additional open-source model (\texttt{DeepSeek-Coder-V2}) 
reported in Section~\ref{sec:rq2}.
We compare our two distance-aware uncertainty metrics, \rao{} and \dcd{},
against three sample-based uncertainty baselines:
DiffTrust~\citep{valentin2026incoherence}, an incoherence-based
behavioural-disagreement metric for code generation;
HonestCoder~\citep{li2024selective}, which combines semantic embeddings and
dataflow features with syntactic n-gram features over sampled programs; and
Semantic Entropy~\citep{farquhar2024semantic}, which clusters
programs based on binary semantic-equivalence judgments from an LLM judge
and computes Shannon entropy over the resulting clusters.

\begin{table*}[h]
\centering
\small
\setlength{\tabcolsep}{4pt}
\renewcommand{\arraystretch}{1.15}
\caption{
AUROC for pass@1 failure prediction across task difficulty levels and
models ($T=0.6$). 
}
\label{tab:auroc_performance_by_difficulty}
\setlength{\tabcolsep}{2pt}
\renewcommand{\arraystretch}{1.10}
\begin{tabularx}{\linewidth}{
>{\raggedright\arraybackslash}X
>{\centering\arraybackslash}p{0.10\linewidth}
>{\centering\arraybackslash}p{0.10\linewidth}
|>{\centering\arraybackslash}p{0.12\linewidth}
>{\centering\arraybackslash}p{0.14\linewidth}
>{\centering\arraybackslash}p{0.10\linewidth}
}
\toprule
 & \textbf{\rao{}} & \textbf{\dcd{}} & \textbf{DiffTrust} & \textbf{HonestCoder} & \textbf{Semantic Entropy} \\
\midrule
\multicolumn{6}{l}{\textbf{GPT-3.5-Turbo-0125}} \\
\midrule
LiveCodeBench All    & 0.827 & \textbf{0.844} & 0.669 & 0.646 & 0.605 \\
LiveCodeBench Easy   & 0.786 & \textbf{0.801} & 0.577 & 0.645 & 0.592 \\
LiveCodeBench Medium & 0.718 & \textbf{0.753} & 0.610 & 0.576 & 0.548 \\
LiveCodeBench Hard   & 0.806 & \textbf{0.831} & 0.568 & 0.570 & 0.550 \\
\midrule
\multicolumn{6}{l}{\textbf{GPT-4o-mini}} \\
\midrule
LiveCodeBench All    & 0.826 & \textbf{0.844} & 0.534 & 0.646 & 0.607 \\
LiveCodeBench Easy   & 0.785 & \textbf{0.796} & 0.565 & 0.645 & 0.594 \\
LiveCodeBench Medium & 0.712 & \textbf{0.752} & 0.504 & 0.576 & 0.548 \\
LiveCodeBench Hard   & 0.801 & \textbf{0.829} & 0.523 & 0.570 & 0.553 \\
\midrule
\multicolumn{6}{l}{\textbf{Gemini-2.5-Flash-Lite}} \\
\midrule
LiveCodeBench All    & 0.794 & \textbf{0.808} & 0.688 & 0.668 & 0.639 \\
LiveCodeBench Easy   & \textbf{0.735} & 0.730 & 0.671 & 0.566 & 0.470 \\
LiveCodeBench Medium & 0.742 & \textbf{0.762} & 0.686 & 0.635 & 0.597 \\
LiveCodeBench Hard   & 0.705 & \textbf{0.737} & 0.523 & 0.590 & 0.645 \\
\midrule
\multicolumn{6}{l}{\textbf{Claude-opus-4.5}} \\
\midrule
LiveCodeBench All    & 0.818 & \textbf{0.825} & 0.606 & 0.739 & 0.687 \\
LiveCodeBench Easy   & \textbf{0.483} & 0.482 & 0.473 & 0.423 & 0.452 \\
LiveCodeBench Medium & 0.723 & \textbf{0.733} & 0.556 & 0.605 & 0.572 \\
LiveCodeBench Hard   & 0.815 & \textbf{0.839} & 0.507 & 0.735 & 0.638 \\
\bottomrule
\end{tabularx}
\end{table*}

\begin{table*}[h]
\centering
\scriptsize
\setlength{\tabcolsep}{3pt}
\renewcommand{\arraystretch}{1.10}
\caption{
Correlation with \tpass{} across task difficulty levels and models
($T=0.6$). Left block: Pearson~$r$. Right block: Spearman~$\rho$.
}
\label{tab:correlation_by_difficulty}
\begin{tabular*}{\linewidth}{@{\extracolsep{\fill}}lccccc|ccccc@{}}
\toprule
 & \multicolumn{5}{c|}{\textbf{Pearson $r$}} & \multicolumn{5}{c}{\textbf{Spearman $\rho$}} \\
\cmidrule(lr){2-6} \cmidrule(lr){7-11}
 & \textbf{\rao{}} & \textbf{\dcd{}} & \textbf{DiffTrust} & \textbf{HC} & \textbf{Semantic Entropy}
 & \textbf{\rao{}} & \textbf{\dcd{}} & \textbf{DiffTrust} & \textbf{HC} & \textbf{Semantic Entropy} \\
\midrule
\multicolumn{11}{l}{\textbf{GPT-3.5-Turbo-0125}}\\
\midrule
LiveCodeBench All    & -0.587 & \textbf{-0.622} & -0.259 & -0.344 & -0.240
                     & -0.581 & \textbf{-0.620} & -0.267 & -0.328 & -0.237 \\
LiveCodeBench Easy   & -0.588 & \textbf{-0.645} & -0.184 & -0.259 & -0.169
                     & -0.549 & \textbf{-0.581} & -0.145 & -0.256 & -0.176 \\
LiveCodeBench Medium & -0.412 & \textbf{-0.467} & -0.153 & -0.185 & -0.112
                     & -0.426 & \textbf{-0.474} & -0.144 & -0.179 & -0.105 \\
LiveCodeBench Hard   & -0.474 & \textbf{-0.516} & -0.063 & -0.205 & -0.144
                     & -0.429 & \textbf{-0.492} & -0.044 & -0.162 & -0.121 \\
\midrule
\multicolumn{11}{l}{\textbf{GPT-4o-mini}}\\
\midrule
LiveCodeBench All    & -0.589 & \textbf{-0.624} & -0.024 & -0.344 & -0.240
                     & -0.584 & \textbf{-0.624} & -0.064 & -0.328 & -0.238 \\
LiveCodeBench Easy   & -0.590 & \textbf{-0.644} & -0.180 & -0.259 & -0.170
                     & -0.547 & \textbf{-0.571} & -0.143 & -0.255 & -0.176 \\
LiveCodeBench Medium & -0.416 & \textbf{-0.474} &  0.013 & -0.185 & -0.110
                     & -0.433 & \textbf{-0.488} & -0.033 & -0.179 & -0.106 \\
LiveCodeBench Hard   & -0.470 & \textbf{-0.515} & -0.018 & -0.205 & -0.142
                     & -0.426 & \textbf{-0.485} & -0.059 & -0.162 & -0.123 \\
\midrule
\multicolumn{11}{l}{\textbf{Gemini-2.5-Flash-Lite}}\\
\midrule
LiveCodeBench All    & -0.494 & \textbf{-0.532} & -0.296 & -0.301 & -0.237
                     & -0.491 & \textbf{-0.512} & -0.298 & -0.329 & -0.242 \\
LiveCodeBench Easy   & -0.414 & \textbf{-0.449} & -0.307 & -0.106 & -0.005
                     & \textbf{-0.314} & -0.308 & -0.215 & -0.134 & -0.033 \\
LiveCodeBench Medium & -0.446 & \textbf{-0.499} & -0.346 & -0.249 & -0.179
                     & -0.414 & \textbf{-0.443} & -0.309 & -0.275 & -0.178 \\
LiveCodeBench Hard   & -0.307 & \textbf{-0.370} &  0.025 & -0.221 & -0.237
                     & -0.308 & \textbf{-0.365} &  0.031 & -0.184 & -0.235 \\
\midrule
\multicolumn{11}{l}{\textbf{Claude-opus-4.5}}\\
\midrule
LiveCodeBench All    & -0.552 & \textbf{-0.611} & -0.139 & -0.424 & -0.275
                     & -0.450 & \textbf{-0.461} & -0.139 & -0.350 & -0.287 \\
LiveCodeBench Easy   &  0.005 &  0.026          & -0.056 & \textbf{ 0.071} & 0.057
                     &  0.024 &  0.025          &  0.024 & \textbf{ 0.052} & 0.048 \\
LiveCodeBench Medium & -0.406 & \textbf{-0.492} & -0.082 & -0.177 & -0.099
                     & -0.253 & \textbf{-0.265} & -0.059 & -0.143 & -0.107 \\
LiveCodeBench Hard   & -0.548 & \textbf{-0.635} &  0.008 & -0.446 & -0.252
                     & -0.531 & \textbf{-0.580} & -0.003 & -0.414 & -0.273 \\
\bottomrule
\end{tabular*}
\end{table*}

\paragraph{Discussion of the Claude--Easy outlier.}
On the LCB~Easy split for Claude-opus-4.5 the AUROC drops markedly relative
to the other splits. This appears to be driven by extreme class imbalance:
Claude achieves a pass@1 success rate of $95.34\%$ on this subset, leaving
very few failure instances. AUROC is equivalent to the Wilcoxon rank-sum
statistic~\citep{hanley1982auc}, whose standard error grows rapidly as the
minority class shrinks, so small changes in the ranking of a handful of
failures produce disproportionate fluctuations in the reported value.

\subsection{Sensitivity to Distance Weights}
\label{sec:distance-discussion}

The crash-aware distance weights $(a,b,c)$ are fixed to $(1, 0.8, 0.6)$
throughout the main paper. To quantify the gap between this fixed scheme
and a per-setting tuned alternative, we additionally consider learning
$(a,b,c)$ via an 80:20 train--test split: the weights are fitted on the
training set and the resulting AUROC is measured on the held-out test set.

The default and learned schemes differ by less than $0.011$ in AUROC across
all settings. On \emph{LiveCodeBench}, only a small subset of inputs (consistently
below 3 out of 10) result in abnormal execution and therefore depend on
$(a,b,c)$, which explains the small gap. In domains where abnormal
executions are more frequent, learning the weights is expected to become
more beneficial; for the regimes considered here, the fixed weighting
scheme is sufficient.

\begin{table*}[h]
\centering
\small
\caption{Fixed default weights $(1.0, 0.8, 0.6)$ vs.\ per-setting learned weights.}
\label{tab:adaptive_weights}
\renewcommand{\arraystretch}{1.15}
\begin{tabular*}{\linewidth}{@{\extracolsep{\fill}}lccccc@{}}
\toprule
Setting & $(a, b, c)$ & Default & Learned & $\Delta$ & Abnormal (/10) \\
\midrule
\multicolumn{6}{l}{\textit{Across models (fixed $T=0.6$)}} \\
\midrule
Claude Opus 4.5        & $(0.90, 0.90, 0.25)$ & 0.8248 & 0.8206 & $-$0.0042 & 2.01 \\
Gemini-2.5-Flash-Lite  & $(0.90, 0.90, 0.90)$ & 0.8075 & 0.7974 & $-$0.0101 & 2.69 \\
GPT-3.5-Turbo          & $(0.97, 0.88, 0.88)$ & 0.8442 & 0.8395 & $-$0.0047 & 2.09 \\
GPT-4o-mini            & $(1.00, 0.80, 0.80)$ & 0.8442 & 0.8524 & $+$0.0082 & 2.13 \\
\midrule
\multicolumn{6}{l}{\textit{Across temperatures (fixed GPT-4o-mini)}} \\
\midrule
GPT-4o-mini ($T=0.2$)  & $(0.51, 0.41, 0.31)$ & 0.7789 & 0.7743 & $-$0.0045 & 2.69 \\
GPT-4o-mini ($T=0.6$)  & $(1.00, 0.80, 0.80)$ & 0.8442 & 0.8524 & $+$0.0082 & 2.13 \\
GPT-4o-mini ($T=0.8$)  & $(0.60, 0.60, 0.60)$ & 0.8530 & 0.8446 & $-$0.0084 & 1.82 \\
\bottomrule
\end{tabular*}
\end{table*}

\subsection{Benchmark Details}
\label{app:datasets}

\emph{LiveCodeBench} (LCB) is our primary benchmark, containing 1{,}055
programming tasks stratified into Easy, Medium, and Hard difficulty
levels with a relatively balanced distribution.
\emph{MBPP} contains 257 entry-level Python function-completion
problems.
\emph{HumanEval-X} comprises 164 hand-written problems
parallel-translated into Python, Java, and C++, and is used to test
cross-language transfer.
\emph{BigCodeBench} consists of 1{,}140 repository-scale tasks
featuring cross-file dependencies, multiple interacting components,
and complex control flow.
Together these benchmarks span a wide range of task difficulty (from
entry-level to repository-scale), programming languages (Python,
Java, C++), and problem styles (function completion, competitive
programming, and realistic library usage).

\begin{table}[h]
\centering
\caption{Statistics of the four benchmarks used in this paper.}
\label{tab:benchmarks}
\small
\begin{tabular}{lrlll}
\toprule
\textbf{Benchmark} & \textbf{\#Tasks} & \textbf{Languages} & \textbf{Task style} & \textbf{Difficulty} \\
\midrule
LiveCodeBench~\citep{livecodebench}        & 1{,}055 & Python              & Competitive programming         & Easy / Medium / Hard \\
MBPP~\citep{MBPP_dataset}                  & 257     & Python              & Function completion             & Entry-level          \\
HumanEval-X~\citep{zheng2023codegeex}      & 164     & Python, Java, C++   & Function completion             & Mixed                \\
BigCodeBench~\citep{zhuo2024bigcodebench}  & 1{,}140 & Python              & Repository-scale, library usage & Hard                 \\
\bottomrule
\end{tabular}
\end{table}

\subsection{Abstention Policy for Practical Deployment}
\label{app:abstention}

To support practical use of uncertainty estimates, we introduce a
simple abstention policy that determines whether to accept or reject
the model's top-ranked generated solution. The policy requires no
additional models or supervision; instead, it operationalises our
distance-aware uncertainty metrics as an actionable decision rule.

Formally, given a programming task $\pi$ and its associated
uncertainty score $U(\pi)$, the abstention decision is defined as
\[
\delta(\pi) =
\begin{cases}
1, & U(\pi) \le \tau \quad \text{(accept)} \\
0, & U(\pi) > \tau \quad \text{(abstain)},
\end{cases}
\]
where $\tau$ is a decision threshold selected via cross-validation on
training data and evaluated on held-out validation folds.

Table~\ref{tab:abstention_pass1_strict} reports abstention
performance for GPT-4o-mini under different operating constraints.
In \emph{safety-oriented} settings, we impose an upper bound on the
false positive rate (FPR), defined as the fraction of incorrect
solutions that are accepted. For each constraint, $\tau$ is chosen
to maximise accuracy under strict correctness (Pass@1 $=1$). Under
a highly conservative constraint (FPR $\le 5\%$), both \rao{} and
\dcd{} achieve strong and comparable performance. Under a more
balanced setting (FPR $\le 20\%$), \dcd{} yields a consistent
accuracy improvement over \rao{}.

\begin{table}[h]
\centering
\small
\caption{Abstention policy performance under strict correctness
(Pass@1 $=1.0$) on \emph{LiveCodeBench} using GPT-4o-mini ($T=0.6$).
Results are shown under two representative operating points:
\emph{Strict Safety} and \emph{Balanced}. Accuracy is maximised
subject to the corresponding FPR constraint. Values are reported as
mean $\pm$ standard deviation over $5$-fold cross-validation.}
\label{tab:abstention_pass1_strict}
\setlength{\tabcolsep}{6pt}
\renewcommand{\arraystretch}{1.1}
\begin{tabular}{lccc}
\toprule
\textbf{Metric} & \textbf{Constraint} & \textbf{Accuracy} & \textbf{FPR} \\
\midrule
\rao{}  & FPR $\leq$ 5\%  & 0.745 $\pm$ 0.095          & 0.064 $\pm$ 0.028 \\
\dcd{}  & FPR $\leq$ 5\%  & 0.745 $\pm$ 0.095          & 0.064 $\pm$ 0.028 \\
\midrule
\rao{}  & FPR $\leq$ 20\% & 0.734 $\pm$ 0.088          & 0.069 $\pm$ 0.028 \\
\dcd{}  & FPR $\leq$ 20\% & \textbf{0.750 $\pm$ 0.095} & 0.068 $\pm$ 0.028 \\
\bottomrule
\end{tabular}
\end{table}

\subsection{Execution Timeout}
\label{app:sensitivity-timeout}

Performance is robust to the choice of per-input execution timeout
(Table~\ref{tab:sensitivity_timeout}). Across all settings, the
proportion of \emph{LiveCodeBench} problems that triggered at least one
timeout error is below $2\%$, indicating that the timeout has only a
minor effect on abnormal-termination rates.

\begin{table*}[h]
\centering
\small
\caption{Effect of the per-input execution timeout on \rao{} and
\dcd{} (\texttt{GPT-4o-mini}, $T=0.6$).}
\label{tab:sensitivity_timeout}
\renewcommand{\arraystretch}{1.15}
\begin{tabularx}{\linewidth}{l*{9}{>{\centering\arraybackslash}X}}
\toprule
& \multicolumn{3}{c}{AUROC}
& \multicolumn{3}{c}{Pearson $r$}
& \multicolumn{3}{c}{Spearman $\rho$} \\
\cmidrule(lr){2-4} \cmidrule(lr){5-7} \cmidrule(lr){8-10}
Timeout (s)
& SE & \rao{} & \dcd{}
& SE & \rao{} & \dcd{}
& SE & \rao{} & \dcd{} \\
\midrule
0.1 & 0.802 & 0.804 & 0.819
    & $-$0.505 & $-$0.540 & $-$0.575
    & $-$0.493 & $-$0.534 & $-$0.573 \\
0.2 & 0.824 & 0.826 & 0.844
    & $-$0.543 & $-$0.589 & $-$0.624
    & $-$0.524 & $-$0.584 & $-$0.624 \\
0.3 & 0.798 & 0.808 & 0.821
    & $-$0.499 & $-$0.550 & $-$0.583
    & $-$0.486 & $-$0.545 & $-$0.580 \\
\bottomrule
\end{tabularx}
\end{table*}
\newpage
\section{Notation}
\label{app:notation}

Table~\ref{tab:notation_summary} lists every symbol used in the paper.
Where the pipeline (\S\ref{sec:pipeline}) and metrics
(\S\ref{sec:metrics}) sections use different surface forms for the same
object, both forms are shown joined by `\(\equiv\)'.

\begin{table*}[h]
\centering
\small
\caption{Notation summary. Defaults from \S\ref{sec:setup} are in parentheses.}
\label{tab:notation_summary}
\setlength{\tabcolsep}{6pt}
\renewcommand{\arraystretch}{1.10}
\begin{tabularx}{\linewidth}{%
  >{\centering\arraybackslash}p{0.22\linewidth}
  >{\raggedright\arraybackslash}X}
\toprule
\textbf{Symbol} & \textbf{Description} \\
\midrule
\multicolumn{2}{l}{\textit{(i) Task and model}} \\
\midrule
$d$                                 & Natural-language task description \\
$\mathrm{Coder}(\cdot)$             & Code-generation model \\
$p(\pi\!\mid\! d)$                  & Conditional distribution over programs given $d$ \\
\midrule
\multicolumn{2}{l}{\textit{(ii) Candidate programs}} \\
\midrule
$K$                                 & Number of candidates per task ($K{=}10$) \\
$\pi$, $\pi_k$, $\pi_1$             & Generic / $k$-th sampled / top-ranked candidate \\
$\boldsymbol{\pi}\equiv\Pi$         & Candidate set $\{\pi_1,\dots,\pi_K\}$ \\
\midrule
\multicolumn{2}{l}{\textit{(iii) Test inputs and execution}} \\
\midrule
$N$                                 & Number of test inputs per task ($N{=}10$) \\
$i_\ell\equiv x_\ell$               & The $\ell$-th test input, $\ell\in\{1,\dots,N\}$ \\
$\mathbf{i}\equiv\mathcal{I}$       & Input set $(i_1,\dots,i_N)$, shared across candidates \\
$\pi(i_\ell)\equiv o_\ell(\pi)$     & Outcome of $\pi$ on $i_\ell$: a normal value or labelled error \\
$\mathbf{o}_k\equiv\sigma_{\mathcal{I}}(\pi_k)$
                                    & Execution signature $\bigl(o_1(\pi_k),\dots,o_N(\pi_k)\bigr)$ \\
\midrule
\multicolumn{2}{l}{\textit{(iv) Behaviour clusters}} \\
\midrule
$M$                                 & Number of clusters \\
$C_i$, $|C_i|$                      & The $i$-th cluster and its size \\
$C(\pi)$                            & Cluster containing program $\pi$ \\
$p_i$                               & Empirical cluster probability $|C_i|/K$, $\sum_i p_i{=}1$ \\
$c^*$                               & Index of cluster containing $\pi_1$, i.e.\ $C_{c^*}=C(\pi_1)$ \\
\midrule
\multicolumn{2}{l}{\textit{(v) Distances and weights}} \\
\midrule
$\delta(\cdot,\cdot)$               & Per-input outcome dissimilarity (Eq.~\eqref{eq:per-input-dist}), $\delta\in[0,1]$ \\
$d_{ij}$                            & Cluster-level distance (Eq.~\eqref{eq:cluster-distance}); symmetric, $d_{ii}{=}0$ \\
$a$                                 & Cost when exactly one outcome is abnormal ($a{=}1$) \\
$b$                                 & Cost when both outcomes abnormal, different errors ($b{=}0.8$) \\
$c$                                 & Cost when both outcomes abnormal, same error ($c{=}0.6$) \\
\midrule
\multicolumn{2}{l}{\textit{(vi) Uncertainty metrics}} \\
\midrule

$\mathrm{SDE}$                      & Semantic Distance Entropy (Eq.~\eqref{eq:sde-def}) \\
$\mathrm{DSDE}$                     & Dominant SDE, anchored at $C_{c^*}$ (Eq.~\eqref{eq:dcd-def}) \\
$U(\pi)$                            & Generic uncertainty score (e.g., \rao{} or \dcd{}) used in abstention \\
\midrule
\multicolumn{2}{l}{\textit{(vii) Experimental settings and targets}} \\
\midrule
$T$                                 & Sampling temperature ($T{=}0.6$) \\
$\tau$                              & Decision threshold of the abstention policy \\
$\mathrm{FPR}$                      & False positive rate (incorrect solutions accepted) \\
$\mathrm{pass@1}$                   & Binary: $\pi_1$ passes all reference tests \\
$\mathrm{partial\_pass@1}$          & Fraction of reference tests passed by $\pi_1$, in $[0,1]$ \\
$\mathrm{AUROC}$                    & Area under ROC for predicting $\mathrm{pass@1}$ failure \\
$r$, $\rho$                         & Pearson and Spearman correlations with $\mathrm{partial\_pass@1}$ \\
\bottomrule
\end{tabularx}
\end{table*}
\newpage
\tcbset{
  codecluster/.style={
    enhanced,
    breakable=false,
    colback=gray!5,
    colframe=black!55,
    boxrule=0.4pt,
    arc=1pt,
    left=5pt, right=5pt, top=2pt, bottom=2pt,
    boxsep=2pt,
    fonttitle=\bfseries\small,
    fontupper=\ttfamily\scriptsize,
    before skip=4pt, after skip=4pt,
    before upper={\setlength{\parskip}{0pt}},
  },
  keychar/.style={
    enhanced,
    breakable=false,
    colback=blue!3,
    colframe=black!55,
    boxrule=0.4pt,
    arc=1pt,
    left=5pt, right=5pt, top=2pt, bottom=2pt,
    boxsep=2pt,
    fonttitle=\bfseries\small,
    fontupper=\footnotesize,
    before skip=4pt, after skip=4pt,
  }
}

\lstdefinestyle{codeclusterlst}{
  basicstyle=\ttfamily\scriptsize,
  aboveskip=0pt,
  belowskip=0pt,
  showstringspaces=false,
  breaklines=true,
  columns=fullflexible,
  keepspaces=true,
}

\section{Concrete Case Study Examples} 
\label{app:case_studies} This appendix provides full task descriptions and representative cluster implementations for the two case studies. Each task is sampled $K=10$ times; programs sharing identical execution outcomes on the shared test inputs are grouped into a single execution-behaviour cluster. The cluster membership for all four tasks is summarised in Table~\ref{tab:app_cluster_membership_caseAB}. To complement the results in RQ1, we present two representative case studies that explicitly isolate the effect of semantic distance between execution clusters. The first case study focuses on \emph{binary correctness} (pass@1), while the second examines \emph{partial correctness} (\tpass{}). In both cases, the tasks induce the same number of execution clusters (i.e., the same count of distinct behaviours); differences in correctness therefore arise solely from variation in the \emph{semantic distance} between clusters rather than from the number of observed behaviours.

\begin{table*}[h]
\centering
\small
\setlength{\tabcolsep}{6pt}
\renewcommand{\arraystretch}{1.15}
\caption{
Execution-behaviour cluster membership and cluster probabilities for the
case-study tasks.
}
\label{tab:app_cluster_membership_caseAB}
\begin{tabularx}{\textwidth}{
  >{\raggedright\arraybackslash}p{0.17\textwidth}
  >{\centering\arraybackslash}p{0.12\textwidth}
  >{\centering\arraybackslash}p{0.18\textwidth}
  >{\raggedright\arraybackslash}X
}
\toprule
\textbf{Task ID} & \textbf{Cluster} & \textbf{Cluster prob.} & \textbf{Member program indices} \\
\midrule
\multicolumn{4}{l}{\textbf{Case A: Binary correctness distinction}} \\
\midrule
\texttt{3367}      & $C_0$ & 0.800 (8/10) & [0, 1, 2, 3, 4, 5, 6, 7] \\
\texttt{3367}      & $C_1$ & 0.200 (2/10) & [8, 9] \\
\addlinespace[2pt]
\texttt{abc332\_b} & $C_0$ & 0.200 (2/10) & [0, 3] \\
\texttt{abc332\_b} & $C_1$ & 0.800 (8/10) & [1, 2, 4, 5, 6, 7, 8, 9] \\
\midrule
\multicolumn{4}{l}{\textbf{Case B: Partial correctness under identical pass@1}} \\
\midrule
\texttt{abc326\_b} & $C_0$ & 0.400 (4/10) & [0, 4, 5, 7] \\
\texttt{abc326\_b} & $C_1$ & 0.600 (6/10) & [1, 2, 3, 6, 8, 9] \\
\addlinespace[2pt]
\texttt{3163}      & $C_0$ & 0.400 (4/10) & [0, 1, 3, 8] \\
\texttt{3163}      & $C_1$ & 0.600 (6/10) & [2, 4, 5, 6, 7, 9] \\
\bottomrule
\end{tabularx}
\end{table*}

\subsection{Case A: Binary Correctness Distinction}
\label{app:case_a}

Case~A contrasts two LiveCodeBench tasks whose execution-behaviour cluster
\emph{structures} are identical (in both tasks the 10 sampled programs split
into two clusters of size 8 and 2), yet whose binary correctness outcomes
differ. The contrast isolates the contribution of inter-cluster semantic
distance: dispersion-based statistics cannot distinguish the two cases, but
the distance-aware metrics can.

\begin{figure}[h]
    \centering
    \includegraphics[width=0.5\linewidth]{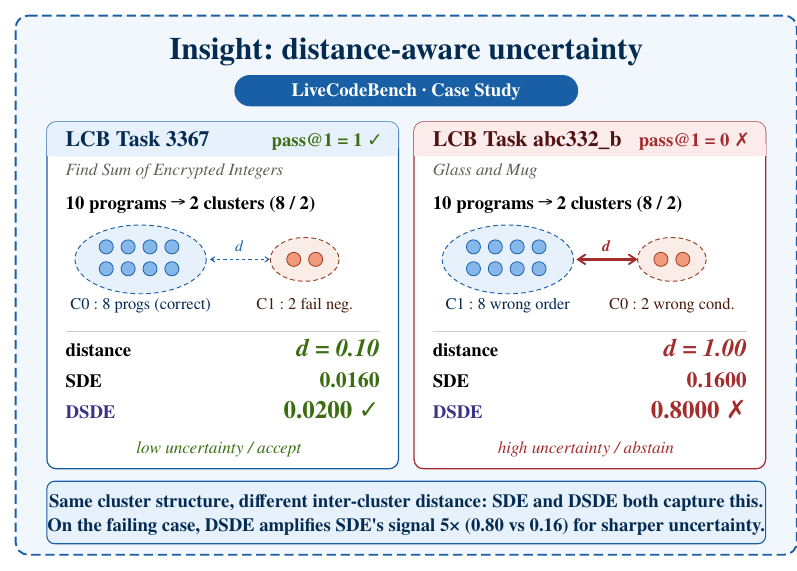}
    \caption{Case~A: two LiveCodeBench tasks with identical execution-behaviour
    cluster structures but different inter-cluster semantic distances and
    binary correctness outcomes.}
    \label{fig:case_a_distance_aware}
\end{figure}

\noindent\textbf{Case A: Semantic Distance Explains Binary Correctness (pass@1).}
Table~\ref{tab:caseA_binary_pass1} compares two tasks with identical cluster
structure ($M=2$, one cluster contains 2 programs, and the other 8 programs)
but different binary outcomes. One task consistently produces
correct executions (pass@1 = 1), while the other fails (pass@1 = 0). Importantly,
both tasks exhibit the same dispersion over execution clusters, ruling out
dispersion-based explanations.

\begin{table}[h]
\centering
\scriptsize
\caption{
Case A: Metric-level comparison illustrating the binary correctness
distinction. Task~1 (\texttt{3367}, $\mathrm{pass@1}=1$) and Task~2
(\texttt{abc332\_b}, $\mathrm{pass@1}=0$) induce an identical $M=2$
cluster structure with sizes 8 and 2, but exhibit very different
inter-cluster distances.
}
\setlength{\tabcolsep}{3pt}
\renewcommand{\arraystretch}{1.10}
\begin{tabularx}{\linewidth}{
p{0.3\linewidth}
>{\centering\arraybackslash}p{0.22\linewidth}
>{\centering\arraybackslash}p{0.22\linewidth}
>{\centering\arraybackslash}p{0.22\linewidth}
}
\toprule
\textbf{Metric} &
\textbf{Task 1 (Low-Distance)} &
\textbf{Task 2 (High-Distance)} &
\textbf{Diff.} \\
\midrule
Pass@1 & 1 & 0 & \textbf{--} \\
\tpass{}  & 1.0000 & 0.0000 & \textbf{1.0000} \\
\midrule
Clusters & 2 (8/2) & 2 (8/2) & \textbf{0}\\
Distance & 0.10 & 1.00 & \textbf{0.90}\\ 
\midrule
\rao{} & 0.0160 & 0.1600 & \textbf{0.1440} \\
\dcd{} & 0.0200 & 0.8000 & \textbf{0.7800} \\
\bottomrule
\end{tabularx}
\label{tab:caseA_binary_pass1}
\end{table}

Although the two tasks induce identical cluster structures, they differ
substantially in the semantic distance between clusters. For the correct task (task 1),
the distance between the two clusters is small (e.g., $d_{01}=0.10$), indicating
highly similar execution outcomes. In contrast, the incorrect task (task 2) exhibits a
much larger inter-cluster distance (e.g., $d_{01}=1.00$), reflecting severe
semantic divergence. This difference in cluster distance directly explains the
observed binary correctness gap and is captured by distance-aware uncertainty
metrics such as \rao{} and \dcd{}.

\subsubsection{Problem 3367: Find the Sum of Encrypted Integers}

\noindent\textbf{Problem statement.}
Given an integer array \texttt{nums}, define \texttt{encrypt(x)} as
replacing every digit in \texttt{x} with the largest digit in \texttt{x}.
Return the sum of encrypted elements. For example,
\texttt{encrypt(523)\,=\,555} and \texttt{encrypt(213)\,=\,333}.

\noindent\textbf{Cluster structure.}
Two clusters with probabilities $p_0 = 0.80$ (8 samples) and
$p_1 = 0.20$ (2 samples). Cluster $C_0$ passes all test cases; cluster
$C_1$ fails on negative-number inputs with a \texttt{ValueError}.

\noindent
\begin{minipage}[t]{0.62\textwidth}
\begin{tcolorbox}[codecluster, equal height group=pair1, title={Cluster $C_0$ — Correct implementation (8 samples)}]
\begin{lstlisting}[style=codeclusterlst, language=Python]
class Solution:
    def sumOfEncryptedInt(self, nums: List[int]) -> int:
        total_sum = 0
        for num in nums:
            max_digit = max(str(num))
            encrypted_num = int(max_digit * len(str(num)))
            total_sum += encrypted_num
        return total_sum
\end{lstlisting}
\end{tcolorbox}
\end{minipage}\hfill
\begin{minipage}[t]{0.36\textwidth}
\begin{tcolorbox}[keychar, equal height group=pair1, title={Key characteristic}]
Uses \texttt{max(str(num))}, which returns the maximum character directly
and handles positive integers correctly.
\end{tcolorbox}
\end{minipage}

\noindent
\begin{minipage}[t]{0.62\textwidth}
\begin{tcolorbox}[codecluster, equal height group=pair2, title={Cluster $C_1$ — Fails on negative numbers (2 samples)}]
\begin{lstlisting}[style=codeclusterlst, language=Python]
class Solution:
    def sumOfEncryptedInt(self, nums: List[int]) -> int:
        total_sum = 0
        for num in nums:
            largest_digit = max([int(digit) for digit in str(num)])
            encrypted_num = int(str(largest_digit) * len(str(num)))
            total_sum += encrypted_num
        return total_sum
\end{lstlisting}
\end{tcolorbox}
\end{minipage}\hfill
\begin{minipage}[t]{0.36\textwidth}
\begin{tcolorbox}[keychar, equal height group=pair2, title={Key characteristic}]
Attempts to convert each character to an integer, raising
\texttt{ValueError: invalid literal for int() with base 10: '-'} on
negative inputs.
\end{tcolorbox}
\end{minipage}

\subsubsection{Problem abc332\_b: Glass and Mug}

\noindent\textbf{Problem statement.}
Given a glass with capacity $G$\,ml and a mug with capacity $M$\,ml, both
initially empty, repeat $K$ operations:
(1)~if the glass is empty, fill the mug completely;
(2)~else if the mug is empty, discard all water from the glass;
(3)~otherwise, transfer water from the mug to the glass (the glass fills
completely if possible, with any remainder staying in the mug).
Output the final water amounts in the glass and the mug.

\noindent\textbf{Cluster structure.}
Two clusters with probabilities $p_0 = 0.20$ (2 samples) and
$p_1 = 0.80$ (8 samples). Both clusters produce incorrect solutions, but
with different error patterns.

\noindent
\begin{minipage}[t]{0.62\textwidth}
\begin{tcolorbox}[codecluster, equal height group=pair3, title={Cluster $C_0$ — Wrong condition check (2 samples)}]
\begin{lstlisting}[style=codeclusterlst, language=Python]
K, G, M = map(int, input().split())
glass, mug = 0, 0
for _ in range(K):
    if glass < G:           # Wrong: should check glass == 0
        glass = G
    elif mug == 0:
        mug = M
    else:
        transfer = min(mug, G - glass)
        glass += transfer
        mug -= transfer
print(glass, mug)
\end{lstlisting}
\end{tcolorbox}
\end{minipage}\hfill
\begin{minipage}[t]{0.36\textwidth}
\begin{tcolorbox}[keychar, equal height group=pair3, title={Key characteristic}]
Uses \texttt{glass < G} instead of \texttt{glass == 0}, misinterpreting
the problem requirements.
\end{tcolorbox}
\end{minipage}

\noindent
\begin{minipage}[t]{0.62\textwidth}
\begin{tcolorbox}[codecluster, equal height group=pair4, title={Cluster $C_1$ — Incorrect operation order (8 samples)}]
\begin{lstlisting}[style=codeclusterlst, language=Python]
K, G, M = map(int, input().split())
glass, mug = 0, 0
for _ in range(K):
    if glass == 0:
        mug = M
    elif mug == 0:
        glass = 0           # Wrong: should fill glass, not empty it
    else:
        transferred = min(M - mug, glass)
        mug += transferred
        glass -= transferred
print(glass, mug)
\end{lstlisting}
\end{tcolorbox}
\end{minipage}\hfill
\begin{minipage}[t]{0.36\textwidth}
\begin{tcolorbox}[keychar, equal height group=pair4, title={Key characteristic}]
The glass is emptied when the mug is empty (line~6), so the glass always
ends at $0$.
\end{tcolorbox}
\end{minipage}

\subsection{Case B: Partial Correctness Distinction}
\label{app:case_b}

Case~B isolates the contribution of inter-cluster distance under
\emph{identical} pass@1 outcomes: both tasks have $\mathrm{pass@1}=0$ and
admit the same 6/4 cluster split, yet differ substantially in
\tpass{}. Distance-aware metrics recover this distinction; dispersion-only
metrics cannot.

\noindent\textbf{Case B: Semantic Distance Explains Partial Correctness.}
Table~\ref{tab:caseB_partial_pass1} presents a complementary scenario,
where both tasks have pass@1 = 0 and induce
identical execution cluster structure ($M=2$, one cluster contains 6 programs, and the other 4 programs), yet differ substantially in
\tpass{}.

\begin{table}[h]
\centering
\scriptsize
\caption{
Case B: Metric-level comparison illustrating partial correctness under
identical binary outcomes. Task~1 (\texttt{abc326\_b}) and Task~2
(\texttt{3163}) both have $\mathrm{pass@1}=0$ and induce an identical
$M=2$ cluster structure with sizes 6 and 4, but differ in their
inter-cluster distance and therefore in \tpass{}.
}
\setlength{\tabcolsep}{3pt}
\renewcommand{\arraystretch}{1.10}
\begin{tabularx}{\linewidth}{
p{0.3\linewidth}
>{\centering\arraybackslash}p{0.22\linewidth}
>{\centering\arraybackslash}p{0.22\linewidth}
>{\centering\arraybackslash}p{0.22\linewidth}
}
\toprule
\textbf{Metric} &
\textbf{Task 1 (Low-Distance)} &
\textbf{Task 2 (High-Distance)} &
\textbf{Diff.} \\
\midrule
pass@1 & 0 & 0 & \textbf{--} \\
\tpass{} & 0.6667 & 0.0000 & \textbf{0.6667} \\
\midrule
Clusters & 2 (6/4) & 2 (6/4) & \textbf{0}\\
Distance & 0.30 & 1.00 & \textbf{0.70}\\ 
\midrule
\rao{} & 0.0720 & 0.2400 & \textbf{0.1680} \\
\dcd{} & 0.1800 & 0.6000 & \textbf{0.4200} \\
\bottomrule
\end{tabularx}
\label{tab:caseB_partial_pass1}
\end{table}

In this case, dispersion-based signals provide no discrimination, as the two
tasks share identical cluster structures and probabilities. However, the
semantic distance between clusters differs markedly. The task with higher
partial correctness exhibits smaller inter-cluster distances (e.g.,
$d_{01}=0.30$), whereas the task with lower partial correctness shows much larger
distances (e.g., $d_{01}=1.00$). These differences align closely with the partial
correctness outcomes and are faithfully reflected by distance-aware metrics.

\subsubsection{Problem abc326\_b: 326-like Numbers}

\noindent\textbf{Problem statement.}
A positive integer is a \emph{326-like number} if it is a three-digit
number satisfying (hundreds digit) $\times$ (tens digit) $=$ (ones digit).
For example, $326$, $400$, and $144$ are 326-like. Given a positive
integer $N$, return the smallest integer $\geq N$ that is a 326-like
number.

\noindent\textbf{Cluster structure.}
Two clusters with probabilities $p_0 = 0.60$
(6 samples, $\mathrm{partial\_pass@1}=1.0$) and $p_1 = 0.40$
(4 samples, $\mathrm{partial\_pass@1}=0.0$). The two clusters differ by
an off-by-one ordering of the increment and the predicate check.

\noindent
\begin{minipage}[t]{0.62\textwidth}
\begin{tcolorbox}[codecluster, equal height group=pair5, title={Cluster $C_0$ — Correct: check before increment (6 samples)}]
\begin{lstlisting}[style=codeclusterlst, language=Python]
N = int(input())
while True:
    hundreds = N // 100
    tens     = (N % 100) // 10
    ones     = N % 10
    if hundreds * tens == ones:   # Check condition first
        print(N)
        break
    N += 1                        # Then increment
\end{lstlisting}
\end{tcolorbox}
\end{minipage}\hfill
\begin{minipage}[t]{0.36\textwidth}
\begin{tcolorbox}[keychar, equal height group=pair5, title={Key characteristic}]
The predicate is checked before incrementing $N$, so the case in which
$N$ itself is already a 326-like number is handled correctly.
\end{tcolorbox}
\end{minipage}

\noindent
\begin{minipage}[t]{0.62\textwidth}
\begin{tcolorbox}[codecluster, equal height group=pair6, title={Cluster $C_1$ — Off-by-one error (4 samples)}]
\begin{lstlisting}[style=codeclusterlst, language=Python]
N = int(input())
while True:
    N += 1                        # Wrong: increment before checking
    hundreds = N // 100
    tens     = (N % 100) // 10
    ones     = N % 10
    if hundreds * tens == ones:   # Then check condition
        print(N)
        break
\end{lstlisting}
\end{tcolorbox}
\end{minipage}\hfill
\begin{minipage}[t]{0.36\textwidth}
\begin{tcolorbox}[keychar, equal height group=pair6, title={Key characteristic}]
$N$ is incremented before the predicate is checked, producing an
off-by-one error whenever $N$ itself is already a 326-like number.
\end{tcolorbox}
\end{minipage}

\subsubsection{Problem 3163: Subarrays Distinct Element Sum of Squares II}

\noindent\textbf{Problem statement.}
Given a $0$-indexed integer array \texttt{nums}, the distinct count of a
subarray is the number of distinct elements it contains. Return the sum
of the squares of the distinct counts of all subarrays. For example,
with \texttt{nums = [1,2,1]}, the answer is
$1^2 + 1^2 + 1^2 + 2^2 + 2^2 + 2^2 = 15$.

\noindent\textbf{Cluster structure.}
Two clusters with probabilities $p_0 = 0.40$
(4 samples, $\mathrm{partial\_pass@1}=0.0$) and $p_1 = 0.60$
(6 samples, $\mathrm{partial\_pass@1}=1.0$). The two clusters differ in
\emph{what} is being squared.

\noindent
\begin{minipage}[t]{0.62\textwidth}
\begin{tcolorbox}[codecluster, equal height group=pair7, title={Cluster $C_0$ — Wrong: sums squared frequencies (4 samples)}]
\begin{lstlisting}[style=codeclusterlst, language=Python]
from collections import Counter
class Solution:
    def sumCounts(self, nums: List[int]) -> int:
        total = 0
        n = len(nums)
        for i in range(n):
            count = Counter()
            for j in range(i, n):
                count[nums[j]] += 1
                # Wrong: sums (frequency^2), not (distinct count)^2
                total += sum(val ** 2 for val in count.values())
        return total
\end{lstlisting}
\end{tcolorbox}
\end{minipage}\hfill
\begin{minipage}[t]{0.36\textwidth}
\begin{tcolorbox}[keychar, equal height group=pair7, title={Key characteristic}]
Computes $\sum_k f_k^2$, where $f_k$ is the frequency of element $k$,
instead of (number of distinct elements)$^2$---a fundamental
misinterpretation of the specification.
\end{tcolorbox}
\end{minipage}

\noindent
\begin{minipage}[t]{0.62\textwidth}
\begin{tcolorbox}[codecluster, equal height group=pair8, title={Cluster $C_1$ — Correct: squares the distinct count (6 samples)}]
\begin{lstlisting}[style=codeclusterlst, language=Python]
class Solution:
    def sumCounts(self, nums: List[int]) -> int:
        ans = 0
        n = len(nums)
        for i in range(n):
            freq = {}
            distinct_count = 0
            for j in range(i, n):
                if nums[j] not in freq:
                    distinct_count += 1
                    freq[nums[j]] = 1
                else:
                    freq[nums[j]] += 1
                ans += distinct_count ** 2  # Correct: (distinct count)^2
        return ans
\end{lstlisting}
\end{tcolorbox}
\end{minipage}\hfill
\begin{minipage}[t]{0.36\textwidth}
\begin{tcolorbox}[keychar, equal height group=pair8, title={Key characteristic}]
Tracks the number of distinct elements and squares this count, matching
the specification.
\end{tcolorbox}
\end{minipage}

Across both cases, correctness differences cannot be explained by cluster counts
or dispersion alone. Instead, they are driven by the magnitude of semantic
distance between execution clusters. These examples provide concrete, intuitive
evidence that the effectiveness observed in Section~\ref{sec:rq1} arises from incorporating
semantic distance, supporting the use of distance-aware uncertainty estimation
for LLM-generated code.


\end{document}